\documentclass[3p,twocolumn]{elsarticle}

\usepackage{amssymb}

\usepackage{subfigure}  

\usepackage{units}

\usepackage{hyperref}

\usepackage{xcolor}

\journal{arXiv}

\begin{document}

\begin{frontmatter}



\title{Solar Kaluza-Klein axion search with NEWS-G}



\author[labelLyon]{Q.~Arnaud}
\author[labelQueens_DMME]{L.~Balogh}
\author[labelGrenoble]{C.~Beaufort}
\author[labelQueens]{A.~Brossard}
\author[labelQueens_DMME]{J.-F.~Caron}
\author[labelQueens]{M.~Chapellier}
\author[labelQueens]{J.-M.~Coquillat}
\author[labelRMC]{E.~C.~Corcoran}
\author[labelQueens]{S.~Crawford}
\author[labelGrenoble]{A.~Dastgheibi-Fard}
\author[labelAlberta]{Y.~Deng}
\author[labelQueens]{K.~Dering}
\author[labelAlberta]{D.~Durnford}
\author[labelAlberta]{C.~Garrah}
\author[labelQueens]{G.~Gerbier}
\author[labelCEA]{I.~Giomataris}
\author[labelQueens]{G.~Giroux}
\author[labelLaurentian,labelSNOLAB,labelMcDonald]{P.~Gorel}
\author[labelCEA]{M.~Gros}
\author[labelQueens]{P.~Gros}
\author[labelGrenoble]{O.~Guillaudin}
\author[labelPNNL]{E.~W.~Hoppe}
\author[labelBirmingham]{I.~Katsioulas}
\author[labelRMC]{F.~Kelly}
\author[labelBirmingham]{P.~Knights}
\author[labelLaurentian,labelSNOLAB,labelMcDonald]{S.~Langrock}
\author[labelSubatech]{P.~Lautridou}
\author[labelQueens]{R.~D.~Martin}
\author[labelCEA]{J.-P.~Mols}
\author[labelGrenoble]{J.-F.~Muraz}
\author[labelBirmingham]{T.~Neep}
\author[labelBirmingham]{K.~Nikolopoulos}
\author[labelAlberta]{P.~O'Brien}
\author[labelAlberta]{M.-C.~Piro}
\author[labelGrenoble]{D.~Santos}
\author[labelQueens]{G.~Savvidis}
\author[labelThessaloniki]{I.~Savvidis}

\author[labelSubatech]{F.~A.~Vazquez~de~Sola~Fernandez\corref{cor1} }

\author[labelQueens]{M.~Vidal}
\author[labelBirmingham]{R.~Ward}
\author[labelGrenoble]{M.~Zampaolo}

\cortext[cor1]{Corresponding author: vazquez@subatech.in2p3.fr}

\address[labelLyon]{IP2I, Université Lyon 1, CNRS/IN2P3, IP2I-Lyon, F-69622 Villeurbanne, France}
\address[labelQueens_DMME]{Department of Mechanical and Materials Engineering, Queen’s University, Kingston, Ontario K7L 3N6, Canada}
\address[labelQueens]{Department of Physics, Engineering Physics \& Astronomy, Queen’s University, Kingston, Ontario K7L 3N6, Canada}
\address[labelGrenoble]{LPSC, Université Grenoble-Alpes, CNRS/IN2P3, 38026 Grenoble, France}
\address[labelRMC]{Chemistry \& Chemical Engineering Department, Royal Military College of Canada, Kingston, Ontario K7K 7B4, Canada}
\address[labelAlberta]{Department of Physics, University of Alberta, Edmonton, Alberta T6G 2R3, Canada}
\address[labelCEA]{IRFU, CEA, Université Paris-Saclay, F-91191 Gif-sur-Yvette, France}
\address[labelLaurentian]{Department of Physics and Astronomy, Laurentian University, Sudbury, Ontario P3E 2C6, Canada}
\address[labelSNOLAB]{SNOLAB, Lively, Ontario, P3Y 1N2, Canada}
\address[labelMcDonald]{Arthur B. McDonald Canadian Astroparticle Physics Research Institute, Queen’s University, Kingston, Ontario K7L 3N6, Canada}
\address[labelSubatech]{SUBATECH, IMT-Atlantique/CNRS-IN2P3/Université de Nantes, 44307 Nantes, France}
\address[labelPNNL]{Pacific Northwest National Laboratory, Richland, WA 99352, USA}
\address[labelBirmingham]{School of Physics and Astronomy, University of Birmingham, Birmingham B15 2TT, United Kingdom}
\address[labelThessaloniki]{Aristotle University of Thessaloniki, Thessaloniki 54124, Greece}

\begin{abstract}

Kaluza-Klein (KK) axions appear in theories with extra dimensions as higher mass, significantly shorter lifetime, excitations of the Peccei-Quinn axion. When produced in the Sun, they would remain gravitationally trapped in the solar system, and their decay to a pair of photons could provide an explanation of the solar corona heating problem. A low-density detector would discriminate such a signal from the background, by identifying the separation of the interaction point of the two photons. The NEWS-G collaboration uses large volume Spherical Proportional Counters, gas-filled metallic spheres with a spherical anode in their centre. After observation of a single axionlike event in a 42 day long run with the SEDINE detector, a $90\%$ C.L. upper limit of $g_{a\gamma\gamma}<\unit[8.99\cdot10^{-13}]{GeV^{-1}}$ is set on the axion-photon coupling for the benchmark of a KK axion density on Earth of $n_{a}=\unit[4.07\cdot10^{13}]{m^{-3}}$ and two extra dimensions of size $R = \unit[1]{eV^{-1}}$.

\end{abstract}




\end{frontmatter}




\section{INTRODUCTION}
\label{Sec:Intro}

The Peccei-Quinn (PQ), or Quantum Chromo-Dynamics (QCD), axion was first proposed to solve the Strong \emph{CP} problem \cite{Peccei1977}. Nowadays, two main families of models, Kim-Shifman-Vainshtein-Zakharov (KSVZ) \cite{Kim1979, Shifman1980} and Dine-Fischler-Srednicki-Zhitnitsky (DFSZ) \cite{Zhitnitskij1980, Dine1981}, describe the characteristics of the pseudo-particle and the PQ symmetry that would give rise to it, with both families agreeing on a non-vanishing axion mass under $\unit[1]{eV}$. This makes it a good dark matter candidate \cite{Preskill1983,Dine1983,Abbott1983}, adding to the interest in the search for the axion in astroparticle physics. Although it remains undiscovered, current constraints set its mass, $m_{PQ}$, between $\unit[10^{-5}]{}$ and $\unit[10^{-2}]{eV}$ \cite{PDG_axion}, which in turn sets a lower bound on the lifetime of its decay into two photons of $\tau_{a \rightarrow \gamma \gamma} > \unit[10^{15}]{Gyr}$, much longer than the age of the universe \cite{GrilliDiCortona2016, DiLuzio2020}.

However, the properties of the PQ axion change in higher-dimensional theories of low-scale quantum gravity \cite{Dvali1999, Dienes2000, Chang2000}. The hierarchy between the gravitational and Planck scale could be explained if $n$ extra compact dimensions exist through which gravity, but not the Standard Model particles, can propagate. In that case, the Planck scale $M_P$ is just an effective coupling, related to the scale of $(n=4)$ dimensional gravity by: $ M_P^2 = 4\pi R^n M_F^{2+n}$ where $R$ is the compactification radius of the extra dimensions and $M_{F}$ is the fundamental quantum-gravity scale. The axion, propagating through these additional dimensions, would obtain a tower of excitations of much higher mass, named Kaluza-Klein (KK) axions, evenly spaced out in mass by a factor of $1/R$. For two additional dimensions and $M_F\sim100\,\mathrm{TeV}$, one obtains $1/R \sim 1\,\mathrm{eV}$ \cite{DiLella2000}. These excitations would have much shorter lifetimes:

\begin{equation}
\tau_{a_{n} \rightarrow \gamma \gamma} = (\frac{m_{PQ}}{m_{a_{n}}})^3 \tau_{a_0 \rightarrow \gamma \gamma} 
\end{equation}

As an example, assuming a coupling to photons of $g_{a\gamma\gamma}=10^{-11}\,\mathrm{GeV}^{-1}$ (corresponding to an axion mass of $m_{PQ}=10^{-1}\,\mathrm{eV}$), a KK axion with $m_{a_{n} } = \unit[10]{keV}$ has a lifetime of $\tau\sim \unit[10]{Gyr}$, $15$ orders of magnitude smaller than a PQ axion, and just under the age of the universe. 

One potential source for these massive KK axions is the Sun, produced either through:

\begin{itemize}
\item Primakoff effect ($\gamma + Ze \rightarrow Ze + a$, where $Ze$ is the Coulomb field due to nuclei and electrons) \cite{Primakoff1951, Raffelt1986};
\item photon coalescence ($\gamma+\gamma \rightarrow a$) \cite{DiLella2003};
\item ABC reactions, standing for: Atomic axio-recombination ($e+I \rightarrow I^{-}+a$) and axio-deexcitation ($I^{*} \rightarrow I+a$), axio-Bremsstrahlung on electrons or ions ($e+(e,I) \rightarrow e+(e,I)+a$), and Compton scattering ($\gamma+e \rightarrow e+a$) \cite{Barth2013,Redondo2013}.
\end{itemize}

Reference \cite{DiLella2003} used the standard solar model to predict production of KK axions through the first two processes, with ABC reactions being negligible for axion models without tree-level coupling to electrons, such as KSVZ. In particular, they demonstrate that a proportion of such heavy axions, mainly created through photon coalescence, will leave the surface of the Sun with speeds under escape velocity, and will remain trapped in closed orbits in the Solar System, accumulating throughout the Sun's lifetime. Simulations of those KK axion orbits determined that the density of trapped axions depend on the distance from the Sun $r$ as $1/r^4$, reaching beyond the Earth's orbit \cite{DiLella2003}.

This model of solar KK axions could additionally resolve the solar corona heating problem, a long-standing physics puzzle dating back to the 1940s \cite{Grotrian1939, Edlen1942}. X-ray measurements of the Sun reveal an unexpectedly high temperature for the Sun's atmosphere, $3000$ times hotter than the surface below it \cite{Parnell2012}. To solve the apparent thermodynamic contradiction, some additional source of energy must dissipate in the corona without affecting the chromosphere. While there are multiple well-developed theories that could explain this phenomenon \cite{Cranmer2015, DeMoortel2015, Sakurai2017}, current observational capabilities cannot directly detect such heating mechanisms \cite{Klimchuk2006, Erdelyi2007, Parnell2012}. An alternative explanation is proposed in Ref.~\cite{DiLella2003}: the corona is heated up by the photons generated from the decays of the trapped axions surrounding the Sun.

\begin{figure}
\centering
  \includegraphics[width=\linewidth]{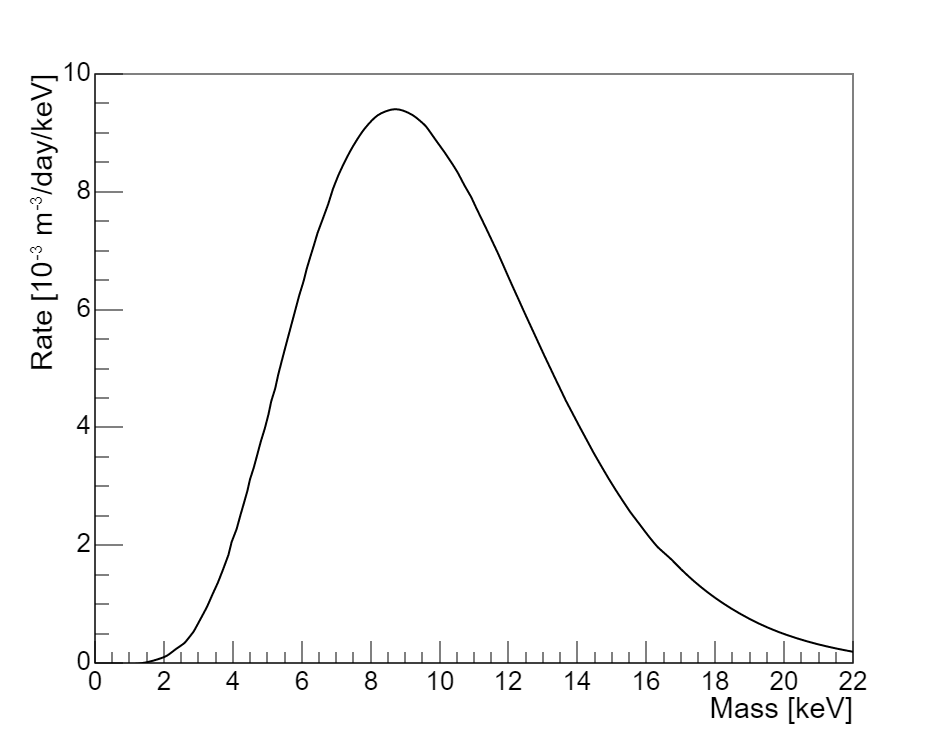}
  \caption{Solar KK axion decay spectrum on Earth (following Ref.~\cite{Morgan2005}, with the density scaling factor described in Ref.~\cite{Oka2017}).}
  \label{fig:AxionDecayDistribution}
\end{figure}

This model would also provide an observable on Earth, in the form of the decays of the local density of KK axions. Figure~\ref{fig:AxionDecayDistribution} shows the expected decay spectrum on Earth. Note that, unlike for decays of an axionlike particle which would appear as a peak with a well-defined energy corresponding to its mass, the solar KK axion model implies a pseudo-continuum of masses, and so predicts a diffuse spectrum of energies. Figure~\ref{fig:AxionDecayDistribution} is scaled for  $g_{a\gamma\gamma} = \unit[9.2\cdot10^{-14}]{GeV}^{-1}$, the predicted axion-photon coupling for which the solar KK axion model can explain the athermal component of the X-ray surface brightness of the quiet Sun, based on ASCA/SIS X-ray data \cite{Orlando2000,Peres2000}. For these values, the local density of axions on Earth is $n_a = \unit[4.07\cdot 10^{13}]{m^{-3}}$, and the integrated decay rate is approximately $\unit[0.08]{event/m^3/day}$, mainly in the $5-\unit[15]{keV}$ range.

The strongest constraint on this model that was described in Ref.~\cite{DiLella2003} comes from limits on exotic energy losses in the Sun derived from core temperature measurements through the solar neutrino flux \cite{Ahmad2002}. An upper bound of $g_{a\gamma\gamma} \sim \unit[10^{-13}]{GeV}^{-1}$ is indirectly set, in mild tension with the preferred value of the axion-photon coupling, at $\unit[9.2\cdot10^{-14}]{GeV}^{-1}$. Given that the model remains relatively unexplored since the original publications, a revision of its predictions, constraints, and dependence on number and size of extra dimensions would be beneficial. In particular, recent measurements of the solar X-ray spectrum of the quiet Sun need to be reanalyzed to update the preferred parameter space of the model, which is likely to reduce the expected decay rate of trapped solar KK axions. However, such a revision falls outside the scope of this work. Until such a revision is carried out, the proposed values of $g_{a\gamma\gamma} < \unit[4.8\cdot 10^{-12}]{GeV}^{-1}$ for $n_a = \unit[4.0\cdot 10^{13}]{m^{-3}}$ should be understood as a benchmark point for the model. At the time of writing, the only constraint derived directly from searches for axion decays on Earth comes from the study of annual modulations in the event rate of the XMASS detector, setting a 90\% confidence level (C.L.) upper limit of $g_{a\gamma\gamma} < \unit[4.8\cdot 10^{-12}]{GeV}^{-1}$ for $n_a = \unit[4.0\cdot 10^{13}]{m^{-3}}$ and two extra dimensions of size $R = \unit[1]{eV^{-1}}$ \cite{Oka2017}, over an order of magnitude above the preferred parameter space.

In this article, a search for solar KK axions using NEWS-G spherical proportional counters is described. In Sec.~\ref{Sec:Detector}, the working principle of such detectors is described, together with the setup of SEDINE, the detector used for this search. Section~\ref{Sec:PulseProcessing} presents the processing of the data to identify axionlike events. Section~\ref{Sec:Simulations} describes the simulations performed to estimate the detector sensitivity to solar KK axions. To validate the approach (both simulations and pulse processing), it was tested on a known source of double-events, $^{55}$Fe-induced argon fluorescence, as covered in Sec.~\ref{Sec:ProofOfConcept}. In Sec.~\ref{Sec:Results}, the extraction of solar KK axion limits from \mbox{SEDINE's} data is explained in detail: first, the calibrations performed to validate the simulations; then the data obtained with \mbox{SEDINE} after a preliminary analysis; finally the use of background simulations to optimize the region of interest for axion searches. Uncertainties and comparison with previous constraints are also described in this section. Finally, the outlook of solar KK axion searches with the upcoming NEWS-G detector at SNOLAB is discussed in Sec.~\ref{Sec:Outlook}.


\section{THE DETECTOR}
\label{Sec:Detector}

\begin{figure}
\centering
  \subfigure {\includegraphics[width=\linewidth]{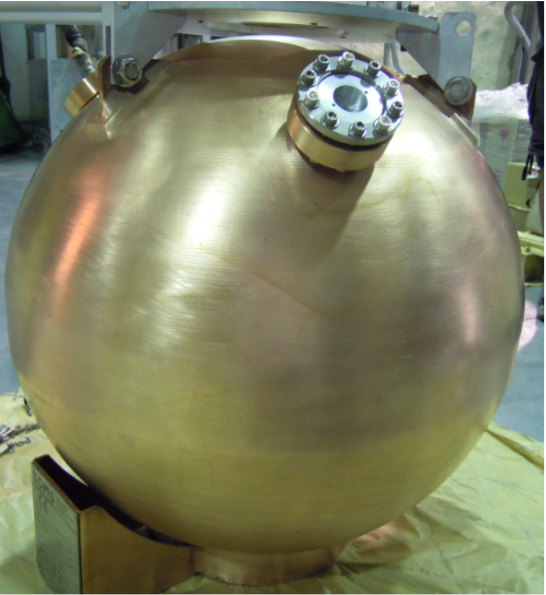}}
  \subfigure {\includegraphics[width=\linewidth]{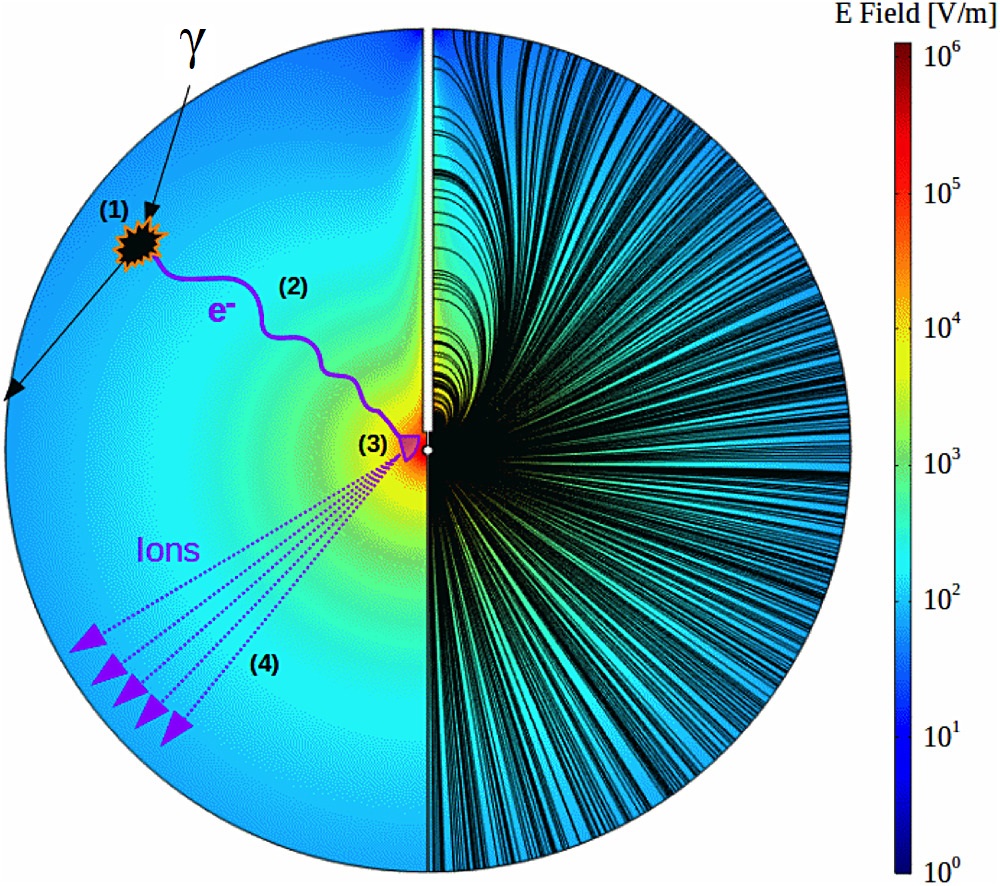}}
  \caption{Top: SEDINE detector (unshielded). Bottom: Working principle of an SPC; (1) Ionization from incident particle (2) Electron drift (3) Avalanche amplification (4) Ion-drift inducing current signal. }
  \label{fig:SPC}
\end{figure}

\subsection{Setup}

SEDINE \cite{Arnaud2018,Brossard2020Thesis}, presented in Fig~\ref{fig:SPC}, is the Spherical Proportional Counter (SPC) \cite{Giomataris2008} installed at the Laboratoire Souterrain de Modane (LSM) in France \cite{Piquemal2012} as a rare event detector for the \mbox{NEWS-G} collaboration. It consists of a $60\,\mathrm{cm}$ inner-diameter grounded spherical shell made of electropure (NOSV) copper \cite{Laubenstein2004} filled with a target gas, and a grounded rod of the same material as the shell supporting a $6.3\,\mathrm{mm}$ diameter silicon anode in the center of the detector; a high voltage is applied on the anode via a wire inside the rod. The shielding of SEDINE comprises three layers: an outer $\unit[30]{cm}$ thick layer of polyethylene against neutrons, a middle $\unit[10]{cm}$ thick layer of lead against $\gamma$ radiation, and an inner $5\,\mathrm{cm}$ thick layer of copper to protect against radiation coming from the lead shield. SEDINE is connected to the outside of the shielding via an $\mathrm{S}$-shaped copper tube, which serves both to connect the central anode to the high voltage source, and the inner volume to the gas handling system.

The data used in this analysis are a $42.7$-day long run using a gas mixture of $99.3\,\%$ neon and $0.7\,\%$ methane at $\unit[3.1]{bar}$, with the central anode at $\unit[2520]{V}$; more details can be found in Ref.~\cite{Arnaud2018}.

\subsection{Working principle}

SPCs are gaseous detectors that record the ionization signal generated by incoming radiation, and can be used in a variety of contexts \cite{Bougamont2017,Savvidis2018,Meregaglia2019}. Their working principle is shown in Fig.~\ref{fig:SPC}. First, an incident particle interacts with the gas in the detector, ionizing atoms which will release primary electrons proportionally to the deposited energy. Thanks to the spherical symmetry of the detector, and ignoring field distortions due to the grounded rod, the electric field in the bulk of the detector is radial and its magnitude scales as $1/r^2$. This field will make the primary electrons drift toward the central anode, diffusing along the way. Their spread in electron arrival times, known as diffusion time and quantified through its standard deviation $\sigma_{PE}$, increases with distance from the central anode as the electrons spend more time drifting; the increased diffusion allows to discriminate against events originating from the inner surface of the shell (e.g. from $^{210}$Pb contamination, see Sec.~\ref{Sec:AdvCuts}) due to the larger width of the signal they induce. Once the drifting electrons reach the intense electric field close to the anode, an avalanche process will release thousands of secondary ion-electron pairs per primary electron, allowing observation of events down to a single primary electron. The secondary ions will induce a current on the anode as they drift away from it, which is then integrated by the readout electronics before being sent to a digitizer.

\subsection{KK axion detection mechanism}

\begin{figure}[t]
\centering
  \subfigure {\includegraphics[width=\linewidth]{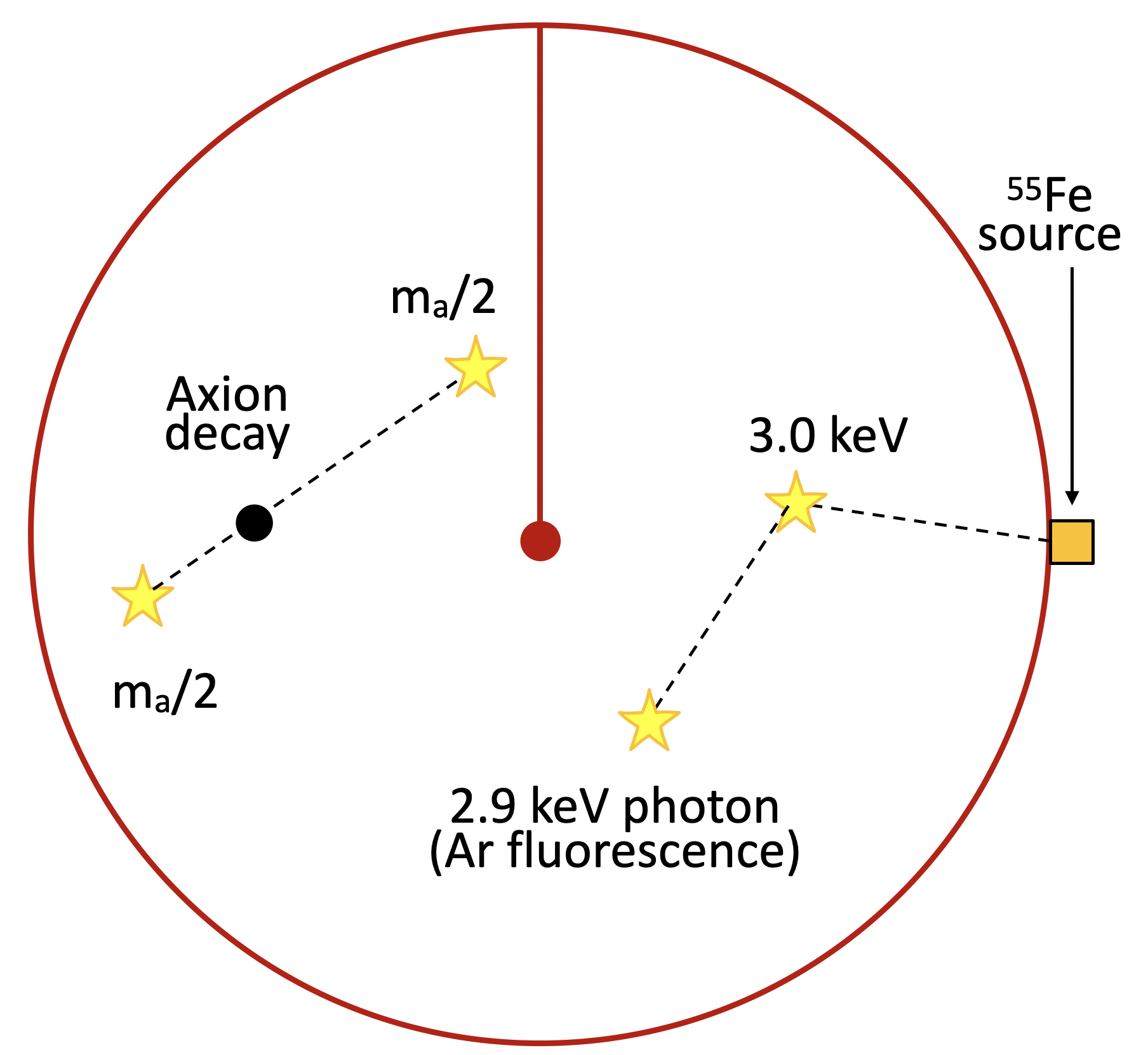}}
  \subfigure {\includegraphics[width=\linewidth]{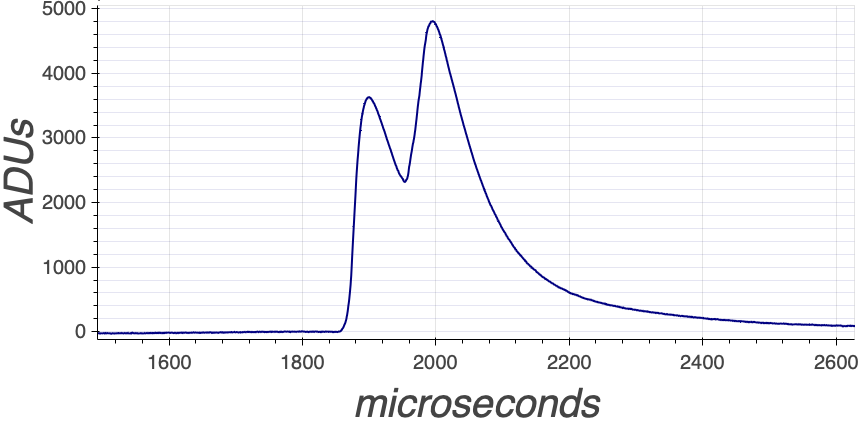}}
  \caption{Top: Schematic of an axion decay inside an SPC; a $^{55}$Fe-induced argon fluorescence event, as described in Sec.~\ref{Sec:ProofOfConcept}, is also shown for comparison.  Bottom: Expected signature of a double-photon event.}
  \label{fig:AxionSignature}
\end{figure}

The use of a low density target mass means an axion decaying in the volume produces two photons that interact at distinct locations, unlike for rare event detectors that use liquids or crystals. Given the increase in the electron drift time with its radial position, axion decays appear as two pulses arriving within a short time interval, as shown in Fig.~\ref{fig:AxionSignature}. This allows for stringent background rejection, by keeping only events with two pulses. Thus, the small amount of active mass, which leads to small exposure in the context of other rare event searches, becomes an advantage for KK axion searches, whose rate depends only on the volume of the detector.

According to the NIST database \cite{NIST, NIST_report},  and in the SEDINE running conditions described in earlier, a $\unit[1]{keV}$, $\unit[4]{keV}$ and $\unit[10]{keV}$ photon have attenuation lengths of $\unit[0.049]{cm}$,  $\unit[2.04]{cm}$ and $\unit[31.6]{cm}$ respectively.  In these conditions, $\unit[2]{keV}$ and $\unit[20]{keV}$ axion decays will not be reconstructed as double pulses, the former due to the short distance between them and the latter due to the likelihood of either or both photons escaping the SPC, but $\unit[8]{keV}$ axion decays will.

\section{PULSE PROCESSING}
\label{Sec:PulseProcessing}

\subsection{Signal formation}

The signal from a single primary electron is the current induced by the ions generated in the avalanche as they drift away from the central sensor. This is given by the Shockley-Ramo theorem \cite{Shockley1938}, which states that a particle induces a portion of its charge on any given electrode given by the electric potential that would exist at the particle's instantaneous position if the selected electrode was at unit potential, all other electrodes at zero potential, and all charges removed \cite{He2001}. Assuming the ideal electric field in $1/r^2$, the theorem can be used to derive the induced current on the anode: 

\begin{equation}
i_{\mathrm{ind}}(t) = -q_{\mathrm{ions}}\alpha\rho(r_{s}^{3}+3\alpha t)^{-\frac{4}{3}} \label{eq:dQ2}
\end{equation}

where $r_{s}$ is the radius of the anode kept at voltage $V_{0}$, $r_{S}$ is the radius of the shell, $1/\rho = 1/r_{s} - 1/r_{S} \simeq 1/r_{s}$, $\alpha=\mu_{0}\frac{V_{0}}{P}\rho$, $\mu_0$ is the ion mobility in the gas, and $P$ its pressure. The current is then fed into a readout charge sensitive preamplifier with response function: 

\begin{equation}
R_{\mathrm{preamp}}(t) = G e^{-t / \tau} \label{eq:preamp}
\end{equation}

where $G$ is the gain of the preamplifier, usually around $\unit[0.2]{V/Me^{-}}$, and $\tau$ is its decay time, in the range of $50-\unit[500]{\mu s}$ for the models used in NEWS-G SPCs. For signals shorter than the decay time, the preamplifier behaves like an integrator, outputting a voltage proportional to the charge generated in the avalanche. The detector impulse response is the convolution of the ion-induced current and the preamplifier response function.

\subsection{Event processing}
\label{Sec:EvtProcessing}

\begin{figure}[t]
\raggedleft
  \subfigure {\includegraphics[width=3.05in]{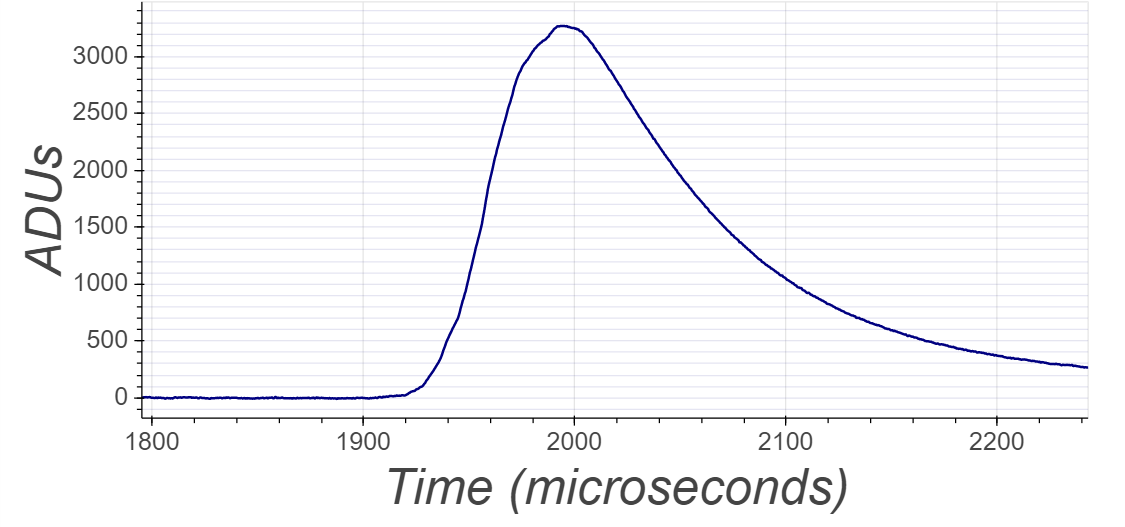}}
  \subfigure {\includegraphics[width=2.97in]{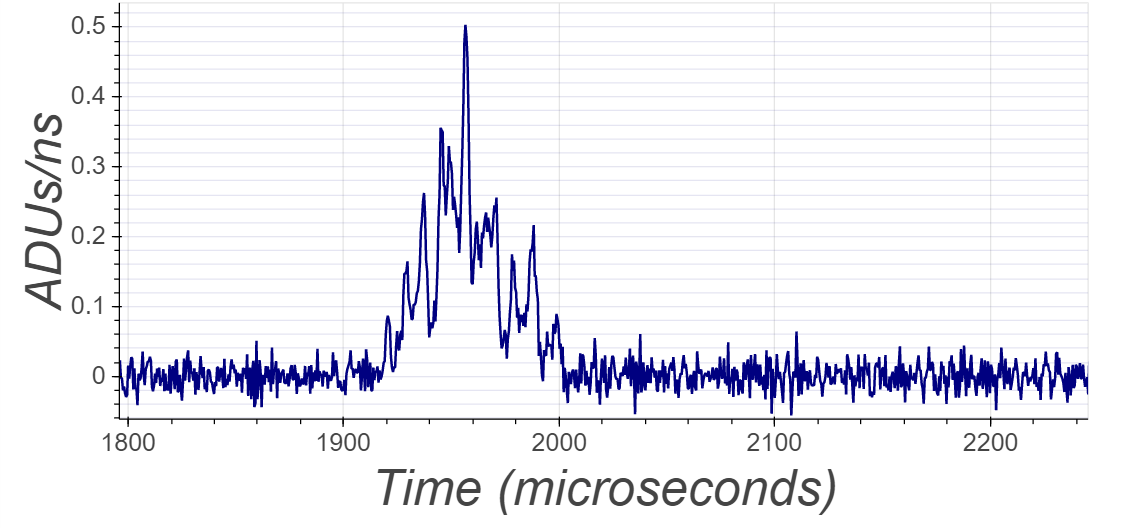}}
  \subfigure {\includegraphics[width=3.1in]{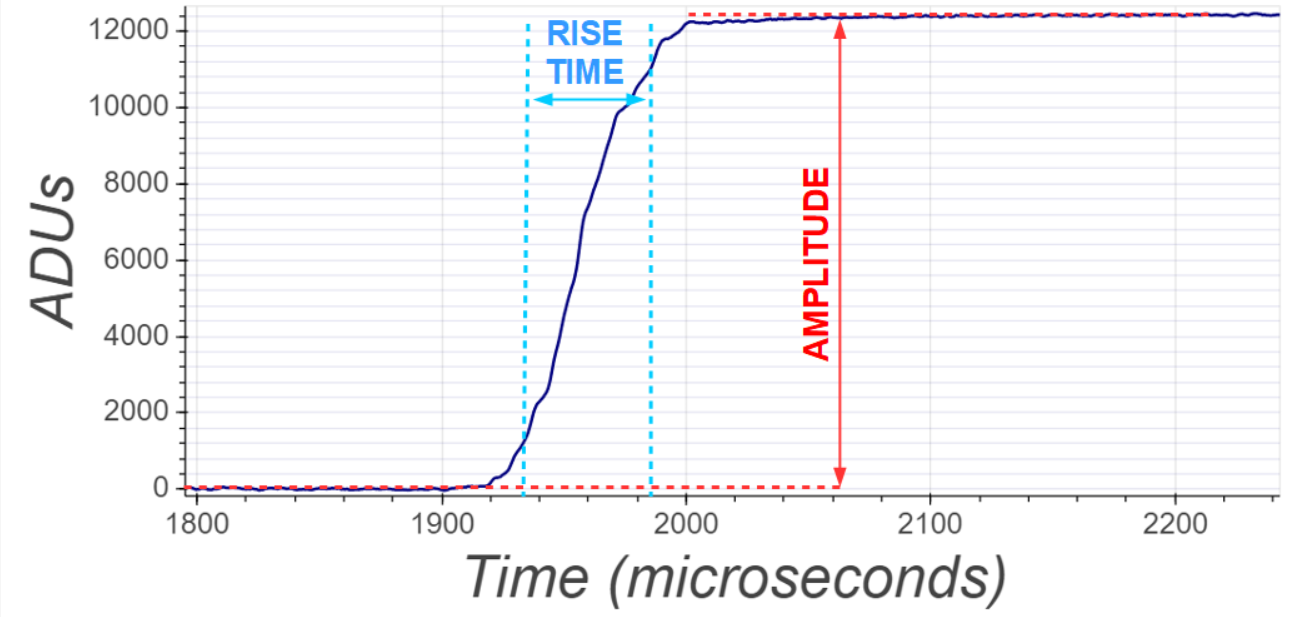}}
  \caption{Example of event producing a single pointlike pulse ($\unit[5.9]{keV}$). Top: Raw signal. Middle: Deconvolved signal; the Gaussian-like structure is driven by the spread in arrival times of the $\sim150$ primary electrons generated by this energy deposition, while the dozen or so spikes are from primary electrons that experienced a particularly high avalanche gain. Bottom: Integrated deconvolved signal; as illustrated, the amplitude and risetime of an event are computed on this signal.}
  \label{fig:PulseProcessing_DD}
\end{figure}

Due to the long drift time of the positive ions with respect to the time constant of the charge-sensitive preamplifier, the recorded signal starts decaying before the whole charge is collected, and so only a fraction of the total charge is measured. This ``ballistic deficit'' can be corrected by deconvolving by the detector impulse response. By definition, the deconvolution of the raw signal results (theoretically) in a signal composed by a series of delta-impulses of varying amplitudes, each corresponding to an avalanche induced by the arrival of a primary electron to the anode. In practice, the use of smoothing over $\unit[5]{\mu s}$ to reduced high-frequency baseline noise imparts a minimum width to the signal originating from individual electrons. For the data used in this work, with $\sim100$ primary electrons per event and a diffusion time under $\unit[20]{\mu s}$, this lead to the individual electron structure being almost completely smeared out. An example of processed event is shown in Fig.~\ref{fig:PulseProcessing_DD}.

Two main parameters are derived from the deconvolved, integrated event. The first is the amplitude, which is directly proportional to the total secondary charge produced in the avalanche, and so in turn to the energy deposited by the incident particle. The second is the risetime, defined as the time it takes for the signal to go from $10\%$ to $90\%$ of the amplitude. The deconvolved pulse is Gaussian-like for a pulse induced by a high number or primary electrons, with the Gaussian's standard deviation driven by their diffusion time, and hence proportional to the risetime of the pulse, $RT = 2.57\,\hat{\sigma}_{PE}$. Since the diffusion time increases with the radial position of the energy deposition, the risetime can be used to discriminate against background events originating from the inner surface of the detector shell. This background rejection technique was already demonstrated for this same dataset in Ref.~\cite{Arnaud2018} in the context of low-mass WIMP-like searches.

\begin{figure}[t]
\raggedleft
  \subfigure {\includegraphics[width=3.05in]{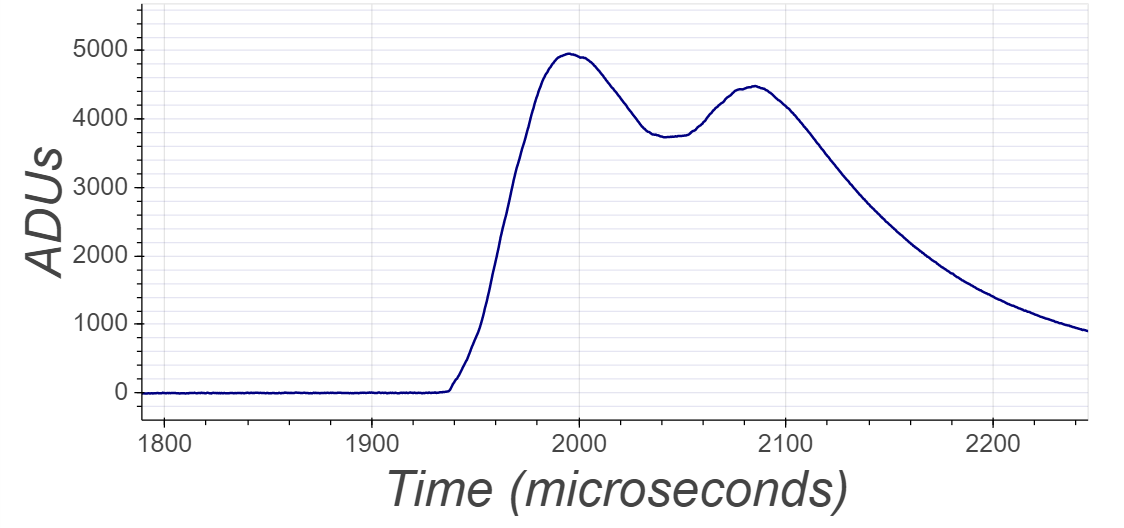}}
  \subfigure {\includegraphics[width=2.97in]{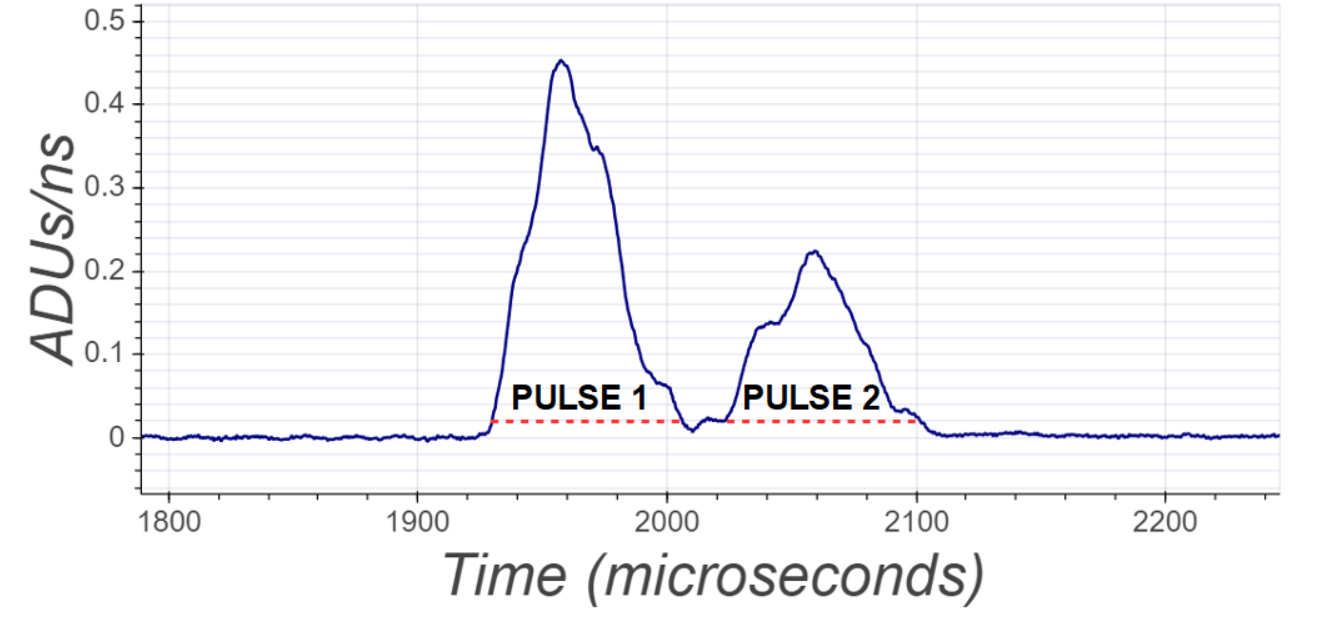}}
  \subfigure {\includegraphics[width=3.1in]{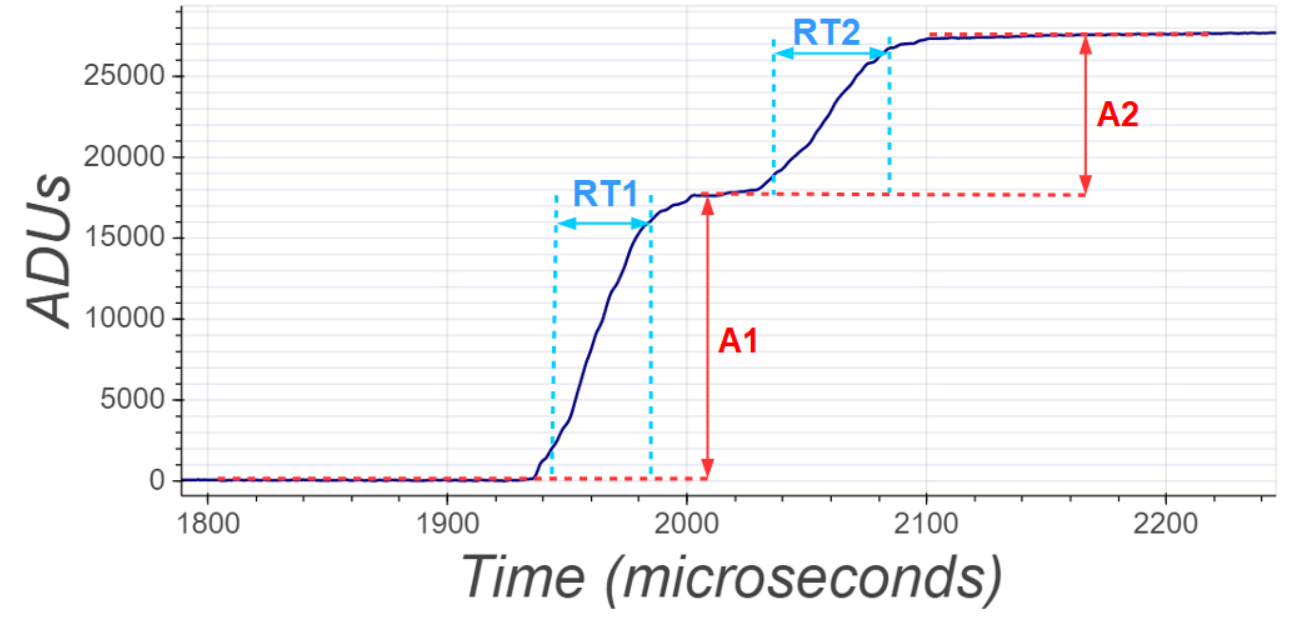}}
  \caption{Example of event producing two pointlike pulses ($\unit[8.5]{keV}$ and $\unit[4.8]{keV}$). Top: Raw signal. Middle: Deconvolved signal, after smoothing over $\unit[20]{\mu s}$; the event is divided into windows in which the signal is above a certain threshold (red dashed line). Bottom: Integrated deconvolved signal; the amplitude and risetime are computed separately in each window; of note, the second pulse has a longer risetime, which is consistent with simultaneous energy depositions at different locations.}
  \label{fig:PulseProcessing_MPA}
\end{figure}

Since axion decays produce two simultaneous energy depositions, one more step is required for their identification. After deconvolving the detector response from the signal, a harsher smoothing over $\unit[20]{\mu s}$ (instead of just $\unit[5]{\mu s}$) is applied, to further reduce baseline noise and clump together the signal from primary electrons close to each other. The event is then divided into windows where the signal stays above a certain threshold. To prevent noise from multiplying the number of threshold crossings when the signal hovers around the threshold, windows closer than 5 samples ($\unit[4.8]{\mu s}$) from each other are merged together, and windows shorter than 5 samples are rejected. This process is demonstrated on Fig.~\ref{fig:PulseProcessing_MPA}, in which it is also apparent how the deconvolution process simplifies the separation of pulses occurring shortly after each other. The threshold and smoothing parameters were chosen so as to optimize the expected limit from the SEDINE data, based on signal and background simulated events (see Sec. ~\ref{Sec:Simulations}). In this, a background of 10k pointlike events were assumed in the $\unit[2-20]{keV}$ energy range, in rough agreement with the data at the stage where only quality cuts were applied \cite{Vazquez2020Thesis}.

The amplitude and risetime of the deconvolved integrated signal (without harsh smoothing) is then computed separately for the pulse in each window. Based only on the characteristics of an axion decay, a preliminary set of requirements was set to identify such events:

\begin{itemize}
\item Event contains two distinct pulses whose amplitudes are above the equivalent of $\unit[500]{eV}$;
\item Both photons from an axion decay have the same energy, so we require the amplitude of both pulses to be the same too, up to the resolution of the detector (22\% at $\unit[2.8]{keV}$); in practice, due to worsening resolution at lower energies and attachment effects (see Sec.~\ref{Sec:Calibrations}), the preliminary version of this cut was deliberately left very wide, taking the form $A_1 < 3 A_2 + \unit[5]{keV}$, where $A_1 > A_2$ are the amplitudes of both pulses;
\item The two interactions happen within $\unit[2]{ns}$ of each other, so the time separation between both pulses cannot be larger than the maximum drift time of primary electrons ($\sim\unit[500]{\mu s}$);
\item The second (i.e. later) pulse comes from primary electrons generated farther from the anode, which experience more diffusion, so the second pulse must have a longer risetime than the first.
\end{itemize}

Additionally, to reduce the number of single-pulse events being falsely reconstructed as containing two pulses, three additional cuts were also added: removing events with a pulse wider than pointlike events observed in the calibration data (cf. Sec.~\ref{Sec:Calibrations}), potentially caused by ``split'' track events; events where the deconvolved signal became significantly negative, indicative of spurious pulses; and events with a saturated signal Tests on simulations (described in the next section) reveal that only the first requirement, the presence of two distinct pulses in an event, has any significant effect on the detection efficiency of axion decays.


\section{SIMULATIONS}
\label{Sec:Simulations}

An ionizing event is simulated in three steps. The first step is to generate the energy deposition within the gas. For pointlike events, such as low energy Compton interactions (Continous Slowing Down Approximation range of a $\unit[10]{keV}$ electron in $\unit[3]{bar}$ of neon is $\unit[0.13]{cm}$ \cite{NIST_STAR}), a position is drawn randomly from the bulk of the detector. For more complex, multi-step interactions, such as an $\alpha$-particle crossing the detector, a complete Geant4 \cite{Agostinelli2003, Allison2006, Allison2016} simulation was used. Each energy deposition is then converted into a number of primary electrons by drawing from a Poisson distribution whose mean is the interaction energy divided by the mean ionization energy $W$, which itself depends on the deposited energy \cite{Inokuti1975}. While measurements of the Fano factor in noble gases are all under $0.20$ \cite{Doke1992}, and estimates for the Fano factor of pure Neon are around $0.13$ \cite{Grosswendt1984,Santos1999}, suggesting that the value for our specific gas mixture should be in that range, it has not been explicitly measured. A value of 1 was assumed, which will lead to a conservative prediction of the detector sensitivity to axion decays, due to assuming a worse energy reconstruction resolution than would be found with a lower Fano factor\footnote{Resolution effects, such as the Fano factor, can have a strong effect on the sensitivity to signals with amplitudes just under the detection threshold \cite{Durnford2018}. However, KK axion decays occur at much higher energies than the $\unit[150]{eV}$ analysis threshold of SEDINE, so this effect will remain minor.}.

The second step is the drift of these primary electrons within the gas due to the electric field in the drift region. This is performed by combining a simulation of the electric field in the detector with COMSOL \cite{COMSOL}, a finite-element analysis software, and Magboltz \cite{Biagi1999}, which computes electron drift velocity and diffusion through Monte Carlo simulations of electron collisions. Using the results from both allows the prediction of primary electron arrival times at the avalanche region.

The third and final step is the generation of the recorded pulse based on those arrival times. Each electron is assigned an amplitude taken from a random draw from a Polya distribution with shape parameter $\theta$, representing the avalanche gain \cite{Alkhazov1970}; in the absence of a $\theta$ measurement in the gas mixture used\footnote{A value of $0.25$, based only on simulations, was used to derive WIMP limits with this same dataset \cite{Arnaud2018}. Since then, laser calibrations have found lower values for $\theta$. Reference \cite{Arnaud2019} in a different SPC found a value of $\theta = 0.09\pm0.02$ in Ne + $2\%$ CH$_{4}$, with even lower values found in unpublished calibrations, motivating the conservative choice of $\theta = 0$.}, since the variance of the Polya distribution is $\sigma_{Polya}^2 = 1/(1+\theta)$, a conservative value of 0 was assumed for the same reasoning previously followed for the Fano factor. Finally, each arrival time and amplitude is then convolved with the response function of the detector described in Sec.~\ref{Sec:Detector}. To include the effect of electronic noise, noise traces recorded with the detector are added onto this ideal pulse.

\begin{figure}[!ht]
\centering
  \subfigure {\includegraphics[width=\linewidth]{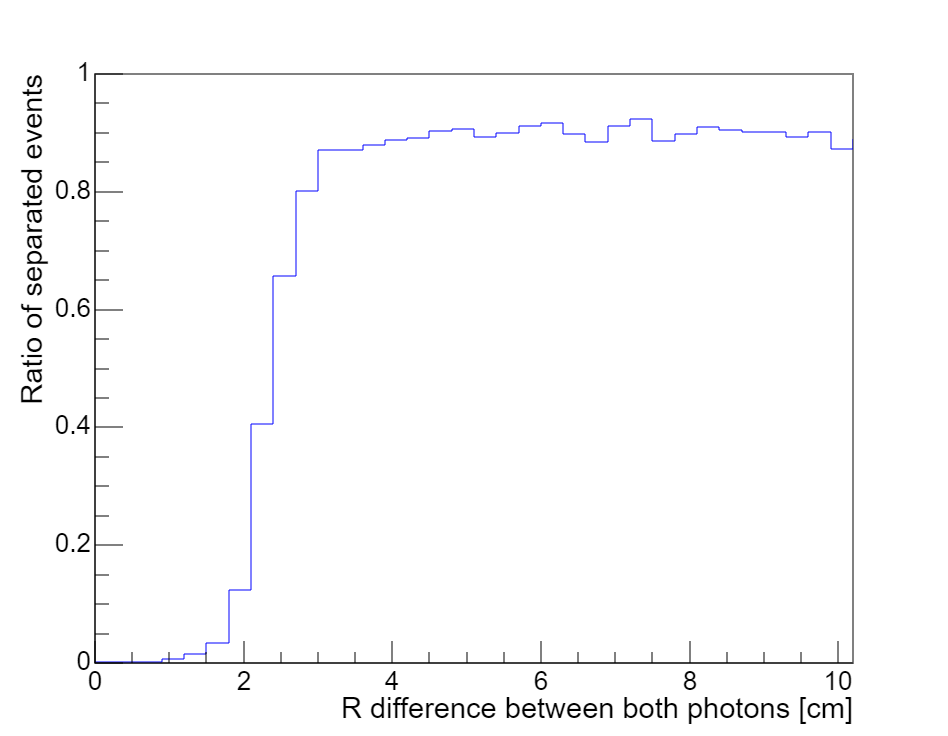}}
  \subfigure {\includegraphics[width=\linewidth]{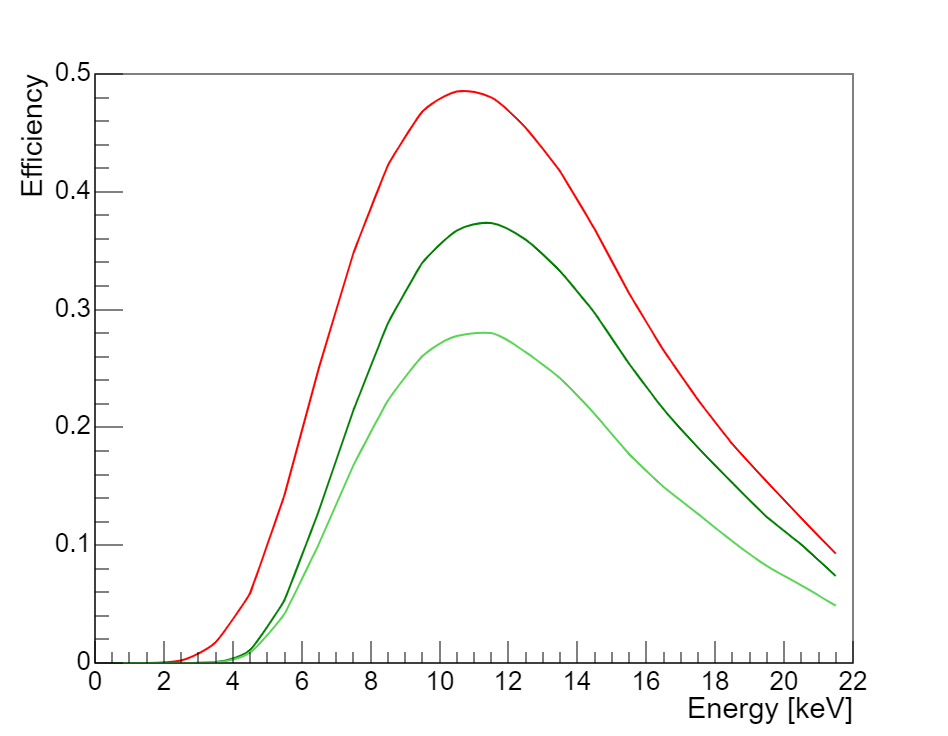}}
  \caption{Results of simulations of the SEDINE detector in the conditions of the physics data. Top: Proportion of axion decay events in which the processing can separate both pulses, depending on the radial difference between the location of the photons' absorptions; the plateau at $90\,\%$ is due to additional pulse shape cuts described in Sec.~\ref{Sec:EvtProcessing}. Bottom: Detection efficiency of solar KK axions. Red: ideal efficiency ignoring smoothing and threshold effects. Dark green: both pulses are separated by processing. Light green: after additional background-rejecting cuts (described in Sec.~\ref{Sec:AdvCuts}).}
  \label{fig:SimEfficiency}
\end{figure}

To simulate solar KK axion events, a random position in the detector is chosen, and two opposite random directions. The mass of the decaying axion was drawn according to their decay distribution (cf. Fig.~\ref{fig:AxionDecayDistribution}), and the distance traveled by the two daughter photons according to that energy and the photon attenuation length as given by the NIST database \cite{NIST, NIST_report}. By processing the simulated axion events with the same algorithm used for the data, the detector efficiency to solar KK axions depending on chosen selection requirements can be computed, as shown in Fig.~\ref{fig:SimEfficiency}. For illustration purposes, in the running conditions of SEDINE,  and for an axion decay at least $\unit[12]{cm}$ away from the center (i.e., $94\,\%$ of SEDINE's volume), the minimum radial separation between both photons before the processing is able to separate both pulses is $\unit[2-3]{cm}$; this range is due to the dependency on the location and energy of the axion decay, both integrated into the final efficiency calculation. Before determination of the optimal selection requirements (described in Sec.~\ref{Sec:AdvCuts}), this leads to an integrated efficiency of SEDINE to solar KK axions of up to $23\%$.

\section{PROOF-OF-CONCEPT WITH $^{55}$FE-INDUCED ARGON FLUORESCENCE}
\label{Sec:ProofOfConcept}

\subsection{Setup}
\label{Sec:PoC_Setup}

To test the multi-pulse analysis methods in preparation for the search of KK axions, a calibration was performed with $^{55}$Fe-induced Argon fluorescence. $^{55}$Fe decays into $^{55}$Mn by electron capture, leaving an electron vacancy in the K-shell, which is then filled by an electron from a higher shell. The difference in energy is then released primarily through two channels, either an Auger electron of $\unit[5.2]{keV}$ (stopped internally in the source) or a K-$\alpha$ X-ray of $\unit[5.9]{keV}$ \cite{TableRadionuclides, TableRadioactiveIsotopes}. In turn, an argon atom that absorbs the $\unit[5.9]{keV}$ photon through the photoelectric effect will release an electron from the K-shell with a binding energy of $\unit[3.2]{keV}$, with the excess $\unit[2.7]{keV}$ transferred as kinetic energy to the electron which dissipates it through further ionizations in the gas; the argon atom, left in an excited state, will fluoresce with a probability of $12\,\%$ (known as the fluorescence yield \cite{Bambynek1972, Krause1979, Hubbell1994}), emitting a photon of $\unit[2.9]{keV}$, leaving the remaining $\unit[0.3]{keV}$ to be dissipated through further low-energy electron or photon emission. At the right pressures, fluorescence will generate two simultaneous and almost identical energy depositions at different positions in the detector: $\unit[2.9]{keV}$ from the fluorescence photon (reabsorbed some distance away from the excited argon atom), and $\unit[3.0]{keV}$ from the photoelectron plus remaining excitation energy (dissipated much closer to it). As shown in Fig.~\ref{fig:AxionSignature}, this is the same signal we would expect from $\unit[5.9]{keV}$ axions decaying within the detector, providing an excellent calibration for multiple-pulse analysis of our signal of interest.

To increase the proportion of such events reconstructed as having two pulses, a low gas pressure is preferred to raise the absorption length of $\unit[3]{keV}$ photons, leading to better separation between both energy depositions. A larger detector was then required to limit the number of photons escaping. The final setup used a $\unit[130]{cm}$ diameter SPC at Queen's University filled with $\unit[200]{mbar}$ of argon with $2\%$ methane, and a high voltage of $\unit[1150]{V}$ applied on a $\unit[2]{mm}$ diameter anode. The pressure and voltage were selected based on Monte Carlo simulations of the detector to maximize the rate of reconstructible fluorescence events. The $\unit[37]{MBq}$ $^{55}$Fe source used was collimated with an aperture of $\unit[1]{mm}$, approximately  $\unit[5]{mm}$ away from the source. The aperture was covered by two sheets of aluminium foil to block $\beta$ radiation, roughly $\unit[20]{\mu m}$ thick each, and placed at the end of a $\unit[4]{cm}$-long window into the detector. A $\unit[213]{nm}$ pulsed laser with a frequency of $\unit[10]{Hz}$ \cite{Arnaud2019} was used concurrently to calibrate for the drift and diffusion time of primary electrons coming from the surface of the detector. Two pulses were considered coincident if they were separated by less than $\unit[250]{\mu s}$, slightly more than the measured surface drift time of electrons ($239\pm\unit[6.3]{\mu s}$) to guarantee no fluorescence event was lost due to this requirement. 

Using a large detector on the surface came with an added difficulty, in the form of cosmic radiation. For its size, an event rate of approximately $\unit[460]{Hz}$ cosmic muons is expected \cite{Shukla2018}. After accounting for the large proportion of pileup due to that high rate, $\unit[579]{Hz}$ of background are indeed observed in the data, compared to only $19.1\pm\unit[1.4]{Hz}$ measured from the  $^{55}$Fe source. Care was taken to separate the true simultaneous events from random coincidences due to the high event rate, as will be shown in the following section.

Furthermore, due to the elevated rate of high-amplitude events, the electric field far from the sensor was dominated by the space charge generated by secondary ions drifting away from the avalanche region, invalidating the electric field computed with COMSOL. However, given the high event rate, a steady-state approximation for the drifting ions can be assumed to derive a new analytical expression for the electric field. For reference, in an ideal spherical symmetry, and in the absence of drifting ions, an application of Gauss's theorem gives $E(r) = \frac{Q}{4\pi\epsilon_0 r^2}$, where $Q\simeq4\pi\epsilon_0 \,r_{s}\,V_{0}$ is the total charge on the anode of radius $r_	{s}$ at potential $V_{0}$. Assuming a steady-state, spherically symmetric, drifting ion space charge, the charge conservation formula leads to a unique solution for the electric field:

\begin{equation}
E(r) = \frac{1}{r^2}\,\sqrt{(\frac{Q'}{4\pi\epsilon_0})^2 + \frac{I_{A}}{6\pi  \epsilon_0 \mu}\,(r^3-r_{s}^3)} \label{eq:EfieldSpaceCharge}
\end{equation}

where $I_A$ is the rate of ion charge creation at the anode (estimated at $\unit[0.19]{nA}$ for these runs), $\mu$ is the ion mobility (estimated at $\unit[1.4]{cm^2/V/\mu s\, bar}$ for this argon mixture), and $Q' \lesssim Q $ is the modified value of the anode charge so that it remains at potential $V_{0}$ and must be computed numerically. This model was used in a Monte Carlo simulation to predict the rate of recognizable fluorescence events, i.e. where the fluorescence photon interacted inside the detector but far enough from the location of the $^{55}$Fe photon absorption. This was found to be $2.8-3.2\%$ of all $^{55}$Fe events. The width of the range of possible values is due to the uncertainty in the minimum radial distance between two simultaneous energy depositions before they can be identified separately, which is  in turn driven by uncertainties in the gas pressure and contaminant presence, and total rate of space charge creation.

\subsection{Results}
\label{Sec:PoC_Results}

\begin{figure}
\centering
\includegraphics[width=\linewidth]{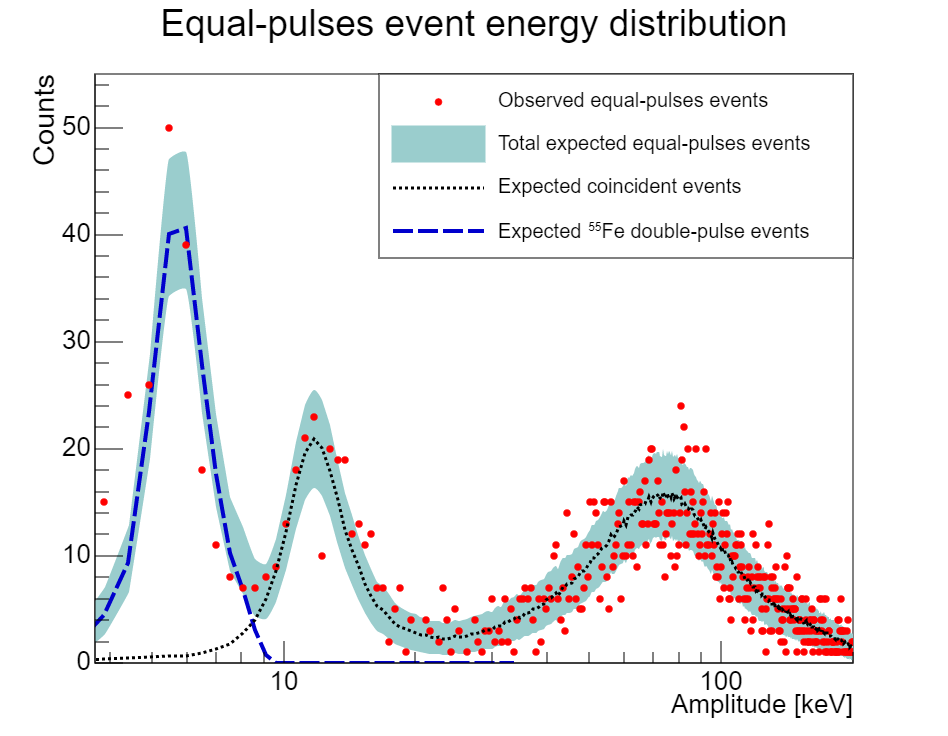}
  \caption{Results of fit on distribution of events with two pulses of same amplitude during the $^{55}$Fe-induced argon fluorescence calibration run. The band around the curve of the total fit represents the statistical uncertainty, which dominates the uncertainties on the fit results. The leftmost peak at $\unit[5.9]{keV}$ corresponds to fluorescence events induced by $^{55}$Fe. The rest of the distribution comes from random coincidences due to the high event rate during the run: the center-left peak at $\unit[12]{keV}$ is due to coincidences of two $^{55}$Fe events, while the diffuse distribution around $\unit[80]{keV}$ is due to coincidences of two cosmic muon events. The contribution from cosmic muons false positives is set to zero by the fit, and so is not shown.}
  \label{fig:ProofOfConcept}
\end{figure}

The energy distribution of all events with coincident pulses of amplitudes consistent with equal energy  is shown in Fig.~\ref{fig:ProofOfConcept}.  The amplitude requirement was set to both amplitudes being within $30\%$ of the average of the two, keeping $>99.9\%$ of all separable fluorescence events, based on the resolution ($8.4\pm2.7\%$) of the $\unit[3.0]{keV}$ escape peak; the uncertainty on this resolution, due to the low statistics of the escape peak and the large background, was the reason for the choice of a wide cut on the amplitude difference. This distribution was fitted by a function with three contributions: recognizable fluorescence events, random coincidences of two single-pulse events, and false-positive events, i.e. single-pulse events falsely reconstructed as being two pulses of equal amplitude. The energy distribution of each of these three contributions was derived directly from the distribution of isolated pulses, which was first split into the $^{55}$Fe peak and all other events \cite{Vazquez2020Thesis}. For this fit, the recognizable fluorescence events followed the same distribution as the main $^{55}$Fe peak, the false-positives followed the distribution of all other events, while the distribution of random coincidences was built by repeatedly taking two random values from the total distribution of isolated pulses, keeping them only if they had equal amplitudes (as defined above). The only parameters in the fit were the number of events of each of these three distributions in the data.

A noticeable result from the fit is how closely the distribution of random coincident events matches both the peak at $12\,\mathrm{keV}$, and the wide population of events around $80\,\mathrm{keV}$, in both shape and relative height. This is despite having no free parameter driving its shape, with the distribution being built directly from the single-pulse data. However, we find $78\%$ more random coincidence events than predicted from a simple $N^2\,\frac{\Delta t}{T}$ formula, with $N$ the total number of events in the run, $\Delta t$ the maximum coincidence time, and $T$ the length of the run. A possible explanation is an overestimation of $T$, most likely due to an underestimation of the effect of pileup on the effective live time.

The False Positive (FP) rate given by the fit is $(0 \pm 0.012)\%$. Given that the fit setting the rate to effectively zero might be tied to the strong assumption that the FP rate is the same at all energies, a different upper limit on the FP rate can be obtained based on the behavior of the distribution around $20\,\mathrm{keV}$, where the fit reaches a local minimum and a statistical uncertainty of $2.7$ events. Since FP events have no discernible effect at that point, and for a total number of events in the run of $\sim150000$, that sets a very conservative upper limit for the FP rate at $0.3\%$ at higher energies. The mismatch between the fit and the data below $\unit[5]{keV}$ comes from false positives being generated at a higher rate at lower energies due to the increased likelihood of electronic noise ``splitting'' a pulse in two, and the worsening resolution weakening the requirement to have both pulses of equal amplitude; this increases the FP rate to $2\%$ under $\unit[5]{keV}$ in these conditions.

Finally, the last term in the fit is the ratio of equal-pulse events in the main $^{55}$Fe peak, with a value of  $2.06\pm0.21\%$ before corrections. The uncertainties in this ratio are dominated by the wide distribution of cosmic muon events reaching down to $\unit[5.9]{keV}$, inducing large uncertainties on the total number of $^{55}$Fe events; uncertainties from contributions of false positives were also taken into account, but had a lesser impact. Furthermore, given an observed total rate of events of  $\unit[600]{Hz}$ and the requirement for coincident events to have exactly two pulses, events with three or more pulses were rejected. This means any fluorescence event had a $34\%$ probability to be rejected due to randomly coinciding with a different event. Correcting for this effect, we get a proportion of equal-pulse events of $(3.1\pm0.3)\%$, in agreement with the $2.8-3.2\%$ from simulations obtained in Sec.~\ref{Sec:PoC_Setup}.

Despite the difficulties brought up by the elevated cosmic background rate, performing this axionlike calibration demonstrated our capability to efficiently identify double-pulse events, the strong rejection of single-pulse events (lower than $0.3\%$ FP rate), and the agreement between our detector model and the data. The first two points prove the suitability of the method to the problem, and the last is a requirement for the extraction of results from real physics data.


\section{KK AXION SEARCH RESULTS} 
\label{Sec:Results}

\subsection{Calibrations}
\label{Sec:Calibrations}

The physics data taken with the SEDINE detector at LSM were calibrated both during the run and with additional measurements.

The risetime of electrons drifting from the surface was calibrated using the radioactive contamination of the inner surface of the detector, which generated a constant background of surface events at all energies during the physics run at a risetime of $50.0\pm\unit[0.6]{\mu s}$. The average time required for electrons to drift from the detector surface was calibrated using the $\beta$ decays from $^{210}$Bi. With a Q value of $\unit[1.16]{MeV}$ \cite{TableRadionuclides, TableRadioactiveIsotopes}, such decays can generate electrons with a CSDA range in 3.1 bars of neon of up to two meters \cite{NIST_STAR}; their range reaches $\unit[30]{cm}$ (SEDINE's radius) starting at $\unit[260]{keV}$. Such high energy electrons cross the detector generating a ``track''' of ionizations along their path. If they start at the inner surface of the detector and pass next to the central electrode, these track events will simultaneously generate electrons next to the sensor (collected almost instantly) and electrons that drift from the surface. Hence, the maximum duration of any track event is the drift time of surface electrons, plus a term that depends on the spread in their arrival times. Risetime calibrations and track simulations allow to remove the diffusion time contribution, obtaining a maximum drift time for electrons of  $422\pm\unit[24]{\mu s}$. Drift time and risetime for events in the gas volume of the detector (rather than the shell's inner surface) were derived using simulations calibrated with the above surface event measurements.

The gain of the detector was estimated with the natural fluorescence of the copper from the detector. We expect a monoenergetic source of fully absorbed X-rays at $8.05\,\mathrm{keV}$ from copper atoms excited by a higher energy $\gamma$. The conversion factor $C$ between the digitizer units (after pulse processing) and energy was found to be $2.08\pm\unit[0.06]{ADU/eV}$; given the electronics gain of $\unit[0.0113]{ADU/e^{-}}$ and a mean ionization energy in neon of $\unit[36]{eV}$ \cite{SauliLectures}, this corresponds to an avalanche gain of $6610\pm190$ secondary ion-electron pairs produced per primary electron; for a Polya distribution with $\theta=0$, this value is also the standard deviation of the avalanche process. Since Ref.~\cite{Arnaud2019} observed a mean ionization energy of neon with $2\%$ of CH$_4$ of $\unit[27.6]{eV}$, compared to the value in the literature for pure neon of $\unit[36]{eV}$, the mean ionization energy was left to vary within the two values as a systematic uncertainty in our analysis, while inversely varying the conversion factor to stay consistent with the observed copper fluorescence. Limiting the possible values of W to this range is conservative, since an even lower value of W for neon with $0.7\%$ of CH$_4$ could only lead to an improved amplitude resolution.

To estimate the effect of using a potentially improper detector response function in the pulse deconvolution step described in Sec.~\ref{Sec:PulseProcessing}, the effective ion mobility during the physics run was computed by deconvolving pointlike pulses with different values of the mobility. The shape of a deconvolved pulse is driven by the arrival time of primary electrons, so pointlike events with enough primary electrons after deconvolution should be a Gaussian driven by the diffusion time during their drift. Hence, the ``optimal'' mobility for each pulse was defined as the one who led to a deconvolved pulse with the best $\chi^2$ after a Gaussian fit. This approach applied to all pointlike events in the $5-\unit[50]{keV}$ range lead to an estimation of the effective ion mobility of $7.45\pm\unit[0.15]{cm^2/V/s}$. This value should not be understood as a measurement of the real ion mobility, but as an \emph{ad hoc} parameter to match the response function of the detector observed in the data. Indeed, it is larger than the value of $\unit[4]{cm^2/V/s}$ for Ne$^{+}$ in pure neon found in the literature \cite{Skullerud1990, Hornbeck1951}, likely due to both the presence of methane both modifying the gas properties and potentially becoming the ion charge carriers, and uncertainties in the electric field close to the anode that are not taken into account in the computation of the effective ion mobility.

Finally, electron attachment during drift to the anode was calibrated by using a gaseous $^{37}$Ar source immediately after the physics run, generating a $2.8\,\mathrm{keV}$ signal uniformly in the bulk of the detector. The probability of a primary electron being attached (primarily due to oxygen contamination) before reaching the sensor increases with its radial distance, since it must drift for longer before being collected. This was observed in the data as a correlation between risetime and amplitude of the $2.8\,\mathrm{keV}$ events. The ratio between the amplitude of high and low risetime events was found to be around $50\%$ on average. However, due to the random nature of electron attachment, high risetime events were observed with amplitudes anywhere between $33\%$ and $100\%$ of the amplitude of low risetime events. Given the low statistics of events at high risetimes, the rate of electron attachment in the simulations was only constrained so that the amplitude ratio between high and low risetime events would remain between $25\%$ and $100\%$.

\subsection{Physics data}
\label{Sec:SEDINEPhysicsData}

During the $42.7$ days of physics data taken with the SEDINE detector, $1639360$ events were recorded.  An offline deadtime of $\unit[2]{s}$ was implemented to remove high-rate periods, observed primarily after high-energy $\alpha$ events, reducing the run length to an effective $\unit[38.0]{days}$. After a preliminary analysis of pulse shapes to reject non-physical (noise transients, spurious pulses, etc.) or track events, and restricting the energy range to $2-\unit[22]{keV}$, $13947$ events are retained. Per Sec.~\ref{Sec:EvtProcessing}, the preliminary region of interest, based only on the expected signal from axions (an adapted event selection that also takes into account our expected background will be described in the next section), was restricted to those with two pulses separated by less than $\unit[500]{\mu s}$, the second one being wider than the first, and where the amplitude of the largest pulse was no larger than thrice the amplitude of the smallest plus $\unit[5]{keV}$. The additional cuts on the shape of the pulses to remove single-pulse events that were improperly split by the processing described in Sec.~\ref{Sec:EvtProcessing} were also applied.

From the $13947$ events in the considered energy range, $284$ contained two ``physical'' pulses, for a combined reduction in background of almost a factor $50$. In turn, applying the fairly loose requirement on the relative amplitudes as the last cut reduced the number of events from $284$ down to $44$, for an additional reduction of over a factor $6$. This is to be contrasted with their effect on the expected axion signal: simulations give an axion detection efficiency of $26.5\%$ when selecting only those leaving two distinct pulses in the detector, and only slightly less at $25.5\%$ when taking all further cuts combined; the same cuts applied on all physical events in the data left only $0.43\%$.

\subsection{Further background rejection}
\label{Sec:AdvCuts}

\begin{figure}[t]
\centering
  \subfigure {\includegraphics[width=\linewidth]{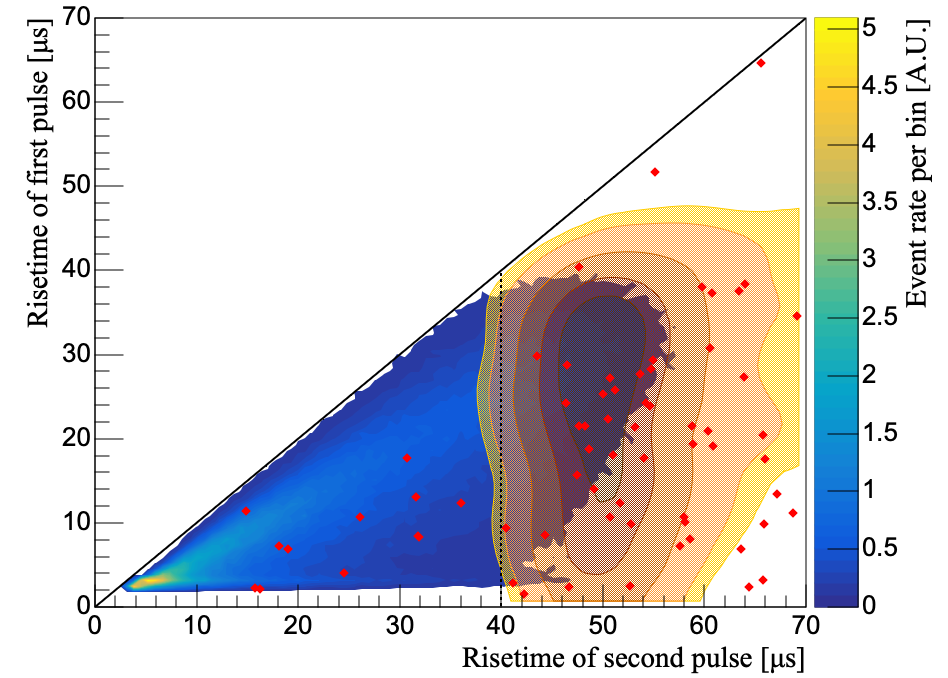}}
  \subfigure {\includegraphics[width=\linewidth]{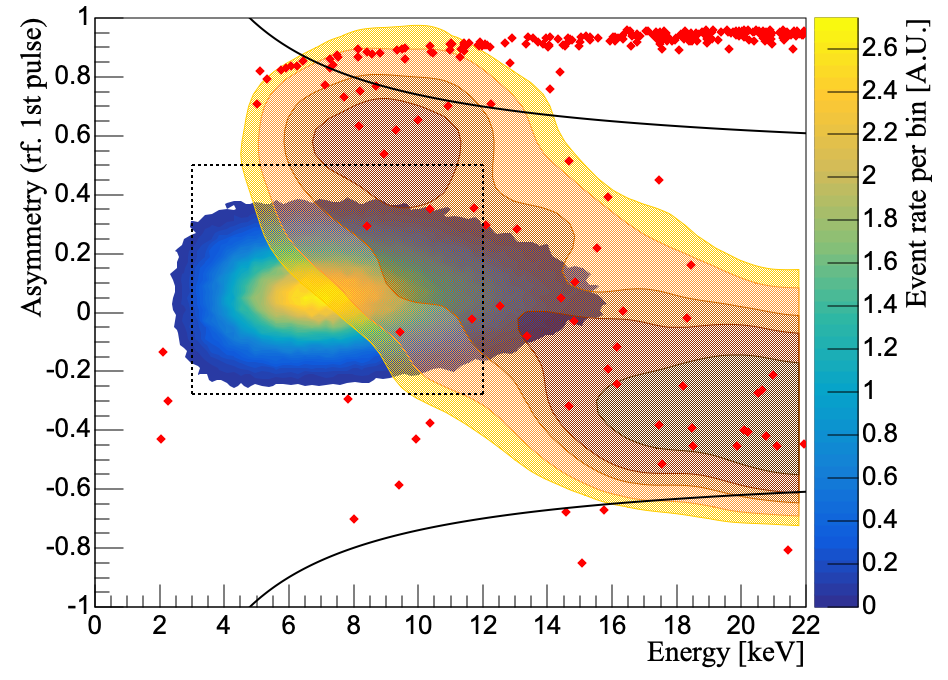}}
  \caption{Comparison between solar KK axion (blue-yellow color plot, with z-axis in arbitrary units) and radioactive background (orange contours,  covering 25, 50, 75, 90 and 95\% of all simulated background) simulations. The solid black lines show the limits of the preliminary region of interest, the dashed lines the limits of the refined one. The physics data (red dots) is also shown for comparison.}
  \label{fig:AxionBackgroundSIM}
\end{figure}

To improve on the results obtained with the preliminary cuts, which were based only on the expected axion signal, the expected background introduced by radioactive contamination in the detector was studied. Geant4 simulations revealed that the primary background was the presence of $^{210}$Pb deposited on the inner surface of the detector shell from the $^{222}$Rn chain, and the second most important contribution was the presence of $^{210}$Bi in the bulk of the shell and lead shield \cite{Brossard2020Thesis}.

\begin{table*}[t]
\resizebox{\linewidth}{!}{
\begin{tabular*}{\textwidth}{lllllll}
Cut                      & \begin{tabular}[c]{@{}l@{}}Physics\\data\end{tabular} & \begin{tabular}[c]{@{}l@{}}Axion\\simulation\end{tabular} & \begin{tabular}[c]{@{}l@{}}Background\\simulation\end{tabular}& \begin{tabular}[c]{@{}l@{}}$^{210}$Pb\\(Cu surface)\end{tabular} & \begin{tabular}[c]{@{}l@{}}$^{210}$Bi\\(Cu bulk)\end{tabular} & \begin{tabular}[c]{@{}l@{}}$^{210}$Bi\\(Pb shield)\end{tabular}  \\
\hline
Quality cuts   & 13078                  & 95.5\%                      & 9050                                  & 3540                           & 2000                        & 1890                        \\
Prel. axion cuts   & 44                     & 25.5\%                      & 111                                   & 83.7                           & 14.2                        & 3.96                        \\
+ adv. risetime cut  & 6                      & 19.2\%                      & 5.06                                  & 1.78                           & 1.00                        & 1.50                        \\
+ adv. asym. cut & 2                      & 19.2\%                      & 4.07                                  & 1.42                           & 0.795                       & 1.16                        \\
+ adv.  energy cut    & 1                      & 17.9\%                      & 2.23                                  & 0.769                          & 0.476                       & 0.476                      
\end{tabular*}
}
\caption{Effect of the preliminary and advanced cuts on data and simulations in the $\unit[2-22]{keV}$ range. The numbers for the data column (resp. background simulation and specific contributions) are the number of observed (resp. expected) events for the $\unit[38.0]{days}$ SEDINE run. Background simulations are taken from Ref.~\cite{Brossard2020Thesis}. The numbers for the axion simulation column is the proportion of simulated axion decays under $\unit[22]{keV}$ that pass the cuts.}
\label{Tab:AdvCuts}
\end{table*}

$^{210}$Pb decays into $^{210}$Bi, which de-excitates by emitting a combination of electrons and photons totaling $\unit[46.5]{keV}$ \cite{Agnese2013}. A photon in the $\sim \unit[10]{keV}$ energy range travels a few centimetres in the gas before being captured \cite{NIST, NIST_report}, while the electrons interact at the location of the decay. Up to $2.3\%$ decays of $^{210}$Pb on the inner surface of the detector that leave energy in the gas were reconstructed as axionlike signals, for a total rate of $\unit[2.2]{events/day}$ from surface contamination in the $\unit[2-22]{keV}$ range. $^{210}$Bi decays by $\beta$-decay, releasing $\unit[1.161]{MeV}$; decays in the bulk of the copper will produce Bremsstrahlung photons, which might do multiple Compton interactions in the gas, or interact at the same time as the original electron. Only about $0.0003\%$ of all $^{210}$Bi decays in the copper bulk that leave energy in the gas are reconstructed as axionlike, for a total rate of $0.37$ events per day in that same energy range. Finally, the fluorescence of copper may generate $\unit[8.05]{keV}$ photons as part of any other energy deposition that interacts with the copper shell, so an excess of equal-amplitude events around $\unit[16.1]{keV}$ is expected, compared to other energies.

The refinement of the region of interest for axion searches was done based on the comparison between these background simulations and solar KK axion simulations, as shown in Fig.~\ref{fig:AxionBackgroundSIM}. It should be noted that, while this was performed based solely on simulations calibrated with single-pulse data analyzed prior to this work \cite{Brossard2020Thesis}, these additional cuts were defined after observing the region of interest.

The first difference between axion and background simulations is the distribution of the risetimes of the second pulse. For axion events, which will be distributed uniformly in the detector volume, all risetimes are roughly equally probable, while for background events dominated by decays of $^{210}$Pb on the surface, the risetime of the second pulse is concentrated around the values for surface events, at $\unit[40]{\mu s}$ and above. Removing all events where the second pulse has a risetime of $\unit[40]{\mu s}$ or more will then remove $95.4\%$ of the total background while keeping $75\%$ of the signal.

The second difference is in the distribution of the asymmetry between the amplitudes of first and second pulse, defined as $(A_1-A_2)/(A_1+A_2)$, where $A_1$, $A_2$ are the amplitudes of the first and second pulse. For KK axion events, the equal energy of both photons would lead to an asymmetry of zero, up to resolution effects and the bias induced by attachment. However, for backgrounds distributed close to uniformly in energy, all values of the asymmetry are roughly equally likely; the exception is that there is an overpopulation of events in which one pulse is produced by copper fluorescence, and so has an energy of $\unit[8.05]{keV}$, inducing a correlation between total energy between both pulses and asymmetry between both. The large uncertainties on attachment reported in Sec. ~\ref{Sec:Calibrations} result in the mean asymmetry of axion events being between $0.00$ and $0.12$, and its standard deviation between $0.10$ and $0.14$. To avoid potentially large systematic uncertainties on our sensitivity to axion decays, the final asymmetry cut was deliberately left fairly wide still. The updated asymmetry cut was set between $-0.28$ and $0.50$ to keep at least $99\%$ of all such events for any value of attachment allowed by calibrations. The final selection of the energy range was chosen after the previous cuts so as to optimize the signal-over-background ratio of axions compared to radioactive background, which was found to be $3-\unit[12]{keV}$.

The results of the improved asymmetry, risetime and energy cuts are gathered in Table~\ref{Tab:AdvCuts}. They have a combined effect of reducing the efficiency of the detector from $25.5\%$ with the basic axionlike cuts, to a total efficiency of $17.9\%$ (cf. Fig.~\ref{fig:SimEfficiency}); the integrated axion event rate becomes $0.0015$ events per day for the benchmark parameters of the model. On the other hand, the expected rate for background events is reduced from $2.9$ events per day to $0.059$ events per day (now dominated by Compton scattering from $^{210}$Bi in the copper bulk and lead shield, due to the risetime cut), for an expected reduction in background of a factor $50$, or a total of $2.23$ background events expected in the dataset. A similar reduction is observed in the data, with only $1$ event in the update region of interest (statistically compatible with $2.23$), compared to $44$ in the preliminary one, for a combined background rejection of $99.99\%$ in the $\unit[2-22]{keV}$ energy range. The comparison of the data with the simulated background (cf. Fig.~\ref{fig:AxionBackgroundSIM}) in both the updated and preliminary region of interest extended to higher energies suggest respectively that the estimation of bulk $^{210}$Bi contamination (dominant in the former) was qualitatively correct, while the surface $^{210}$Pb contamination (dominant in the latter) might have been overestimated by a factor $3-6$; no retroactive correction of the expected background rate was performed, to avoid biasing the choice of region of interest.

\subsection{Exclusion limit}
\label{Sec:ExclLimit}

\begin{figure}
\centering
  \includegraphics[width=\linewidth]{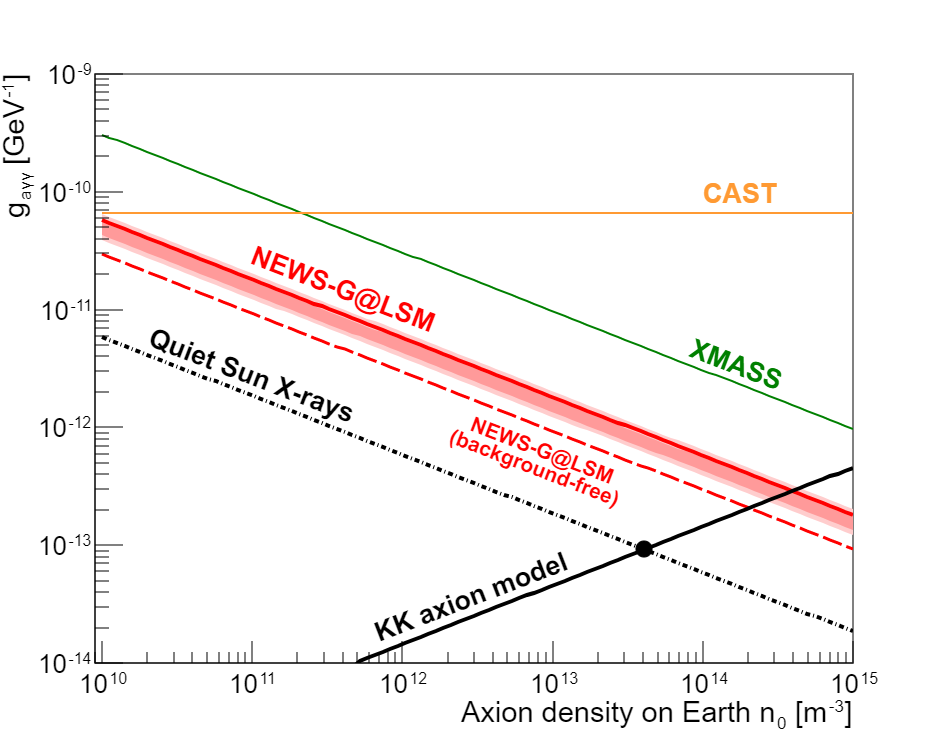}
  \caption{Exclusion limit for solar KK axions derived from this work (solid red line). The $\pm 1,2\,\sigma$ range of values covered by the systematic uncertainties are represented by the red shaded areas. For comparison, we show the ideal exclusion limit in the absence of background and systematics (dashed red line), the previous limits on solar KK axions from CAST (orange line \cite{Anastassopoulos2017}), and from the XMASS collaboration (green line \cite{Oka2017}). The preferred parameter space for the solar KK axion model is shown as the intersection between the solid black line (solar KK axion model) and the dashed black line (quiet Sun athermal X-ray hint) \cite{DiLella2003}.}
  \label{fig:ExclusionLimit}
\end{figure}

A single candidate event was observed in the data after applying all selection criteria described in the previous sections. This is consistent with the expectation of $2.23$ background events.  Together with the expected solar KK axion decay rate, an upper limit is set on the axion-photon coupling:

\begin{equation}
\label{eq:ExclLimit}
g_{a\gamma\gamma_{\mathrm{excl}}}(n_0) = g_{_{\mathrm{DLZ}}} \sqrt{\frac{N_{\mathrm{excl}}} {N_{\mathrm{exp}}}\,\frac{n_{_{\mathrm{DLZ}}}}{n_0}}
\end{equation}

where $N_{\mathrm{excl}}$ is the excluded number of events for our observed number (assuming Poisson statistics \cite{GARWOOD1936} and taking a $90\%$ C.L. upper limit,  $3.89$ given an observation of a single candidate event), $N_{\mathrm{exp}}$ is the expected number of KK axions in the detector given our exposure and efficiency, $g_{a\gamma\gamma}$ is the coupling between axions and photons, $n_0$ is the local density of trapped solar KK axions on Earth,  and $g_{_{\mathrm{DLZ}}}$ and $n_{_{\mathrm{DLZ}}}$ are their predicted values in Ref.~\cite{DiLella2003} to match the X-ray surface brightness of the quiet Sun. The final exclusion limit plot is shown in Fig.~\ref{fig:ExclusionLimit}: at $n_0=n_{_{\mathrm{DLZ}}}=\unit[4.07\cdot10^{13}]{m^{-3}}$, and given $g_{_{\mathrm{DLZ}}}=\unit[9.2\cdot10^{-14}]{GeV^{-1}}$, we obtain an exclusion limit of $g_{a\gamma\gamma}=\unit[8.99\cdot10^{-13}]{GeV^{-1}}$. Compared to the only other existing exclusion limit from axion decays on Earth, set by the XMASS collaboration at $g_{a\gamma\gamma}=\unit[4.8\cdot10^{-12}]{GeV^{-1}}$\cite{Oka2017} for $n_{_{\mathrm{DLZ}}}$,  NEWS-G sets a limit $5.2$ times lower.  At that same KK axion density on Earth, NEWS-G's limit is $70$ stronger than CAST's constraint on axionlike particles generated by the Sun.

The uncertainty on the final exclusion limit was obtained by simulating the effect of calibration uncertainties on the detector sensitivity to axions. 1000 simulations with $\unit[10]{k}$ solar KK axion events each were generated, where the simulation parameters were left to vary freely within the values allowed by calibration uncertainties. The parameters considered were the uncertainties on the calibration of the electron drift and diffusion time, electron attachment, photon attenuation length, energy and mean ionization energy calibrations, and ion mobility in the gas. For the chosen optimized selection criteria, the combined effect from all parameters lead to a relative uncertainty on the total efficiency of $24\%$, which contributes a relative $13\%$ uncertainty on the limit.  The final reported value was taken at the upper $90\%$ point of the range of possible values due to these uncertainties. The main contributions were the diffusion time, with the efficiency acquiring a standard deviation of $13\%$ of its mean when it is the only parameter left to vary, and the drift time, at $11\%$. These large contributions are due respectively to the strengthened risetime cut, and to small differences in drift time affecting the separability of close energy depositions. Due to integrating the calibration uncertainties for attachment in our choice of asymmetry and energy cuts, attachment only induced an uncertainty on our detector efficiency of $5\%$. All other calibrations had individual contributions to the final uncertainty of less than $2\%$.  

For a model independent plot representing the results of this search and how to exploit it, see \ref{App:GenericLimit}.


\section{PROJECTIONS WITH UPCOMING DETECTOR}
\label{Sec:Outlook}

\subsection{Setup}

The next phase for the NEWS-G collaboration is a high purity copper (C10100) $\unit[140]{cm}$ diameter detector  (hereafter S140 detector) at SNOLAB, one of the deepest low-background laboratories in the World \cite{Duncan2010}. A new kind of multi-anode sensor, the ACHINOS \cite{Giganon2017,Giomataris2020}, was developed to accommodate for the larger detector size. Notably, this new type of sensor has the potential for 3D event reconstruction through the use of an individual readout channel for each anode, each pointing in a different direction, leading to increased detector efficiency and background rejection. However, since the performance of such a setup has not been characterized yet, this capability was not taken into account for the present study. $\unit[500]{\mu m}$ of pure copper was electroplated on the inside surface of the detector shell, to attenuate the backgrounds from the $^{210}$Pb contamination on the internal surface of the detector \cite{Balogh2021}. The detector will be enclosed in $\unit[25]{cm}$ of lead, of which the internal $\unit[3]{cm}$ of this shield are made of archaeological lead, and $\unit[40]{cm}$ thick of polyethylene. With improved radiopurity, increased size, gas purification, and continuous calibrations via laser, this new detector should have drastically enhanced sensitivity to solar KK axions. At time of writing, installation is being finalized at SNOLAB. More details on the setup can be found in \cite{Giroux2019}, with a dedicated paper in preparation at the time of writing.

\begin{figure}
\centering
  \includegraphics[width=\linewidth]{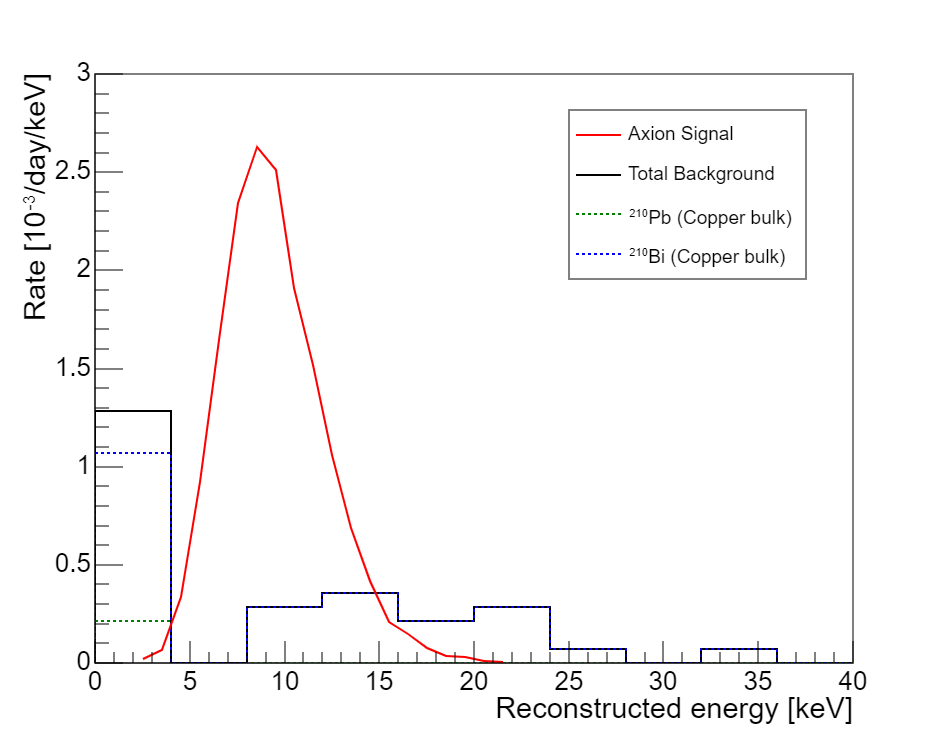}
  \caption{Results from simulations of axions and background in the S140 after analysis cuts, based only on $^{210}$Bi contamination in the copper bulk. Background events under $\unit[6]{keV}$ come primarily from improper reconstruction of low energy events. The rest come from double Compton interactions of Bremsstrahlung photons from $^{210}$Bi, or one Compton interaction together with a copper fluorescence photon.}
  \label{fig:SNOGLOBE_Background}
\end{figure}

S140's background is expected to be considerably lower than SEDINE's. Notably, the electroplating should drastically reduce $^{210}$Pb contamination on the inner surface of the detector, so $^{210}$Bi contamination in the bulk of the copper shell becomes the largest contribution to axionlike background. The activity of $^{210}$Po daughters was measured in a sample of the C10100 copper used to build the S140; the activity of $^{210}$Pb (and hence $^{210}$Bi) was found to be $\unit[29^{+8+9}_{-8-3}]{mBq}$ per kilogram \cite{Balogh2021}. Using the most likely value, and for a total copper mass of $\unit[521.4]{kg}$, a Geant4 simulation with $3\cdot 10^9$ decays (corresponding to $2336\,\mathrm{days}$ of exposure) was performed. The operating conditions used in the simulation were $0.6\,\mathrm{bar}$ of neon with $\unit[2000]{V}$ applied on the anode, which were found to lead to the highest rate of detected axion decays. The comparison between expected axion signal and background is shown in Fig.~\ref{fig:SNOGLOBE_Background}.

In these conditions, and keeping only events in the $\unit[5-15]{keV}$ range, the expected axion event rate with the parameters from Ref.~\cite{DiLella2003} is $\unit[16.5\cdot 10^{-3}]{events/day}$. Comparatively, even with a conservative estimate of a uniform background of $\unit[4\cdot 10^{-4}]{evt/day/keV}$ in the same energy range, the total expected background rate is $\unit[4\cdot 10^{-3}]{events/day}$, roughly four times lower than the axion rate. This is a considerable improvement over SEDINE, for which the expected background rate after event selection ($\unit[59\cdot 10^{-3}]{events/day}$) was forty times higher than the axion rate ($\unit[1.5\cdot 10^{-3}]{events/day}$). Based on this background-to-axion ratio, following the same approach as in Sec.~\ref{Sec:ExclLimit} would require a run length of around $\unit[400]{days}$ to be able to reject at 90\% C.L. the solar KK axion explanation for the athermal X-ray spectrum of the quiet Sun as posited in \cite{DiLella2003}, although with the caveats noted in Sec.~\ref{Sec:Intro} on the need to update both the model and the constraints from the quiet Sun spectrum. For reference, an ideal S140 with $<1\%$ background-to-axion ratio would only require $\unit[180]{days}$ to reach the same sensitivity. Conversely, interpreting the athermal X-rays from the quiet Sun as a solar KK axion signal (with the same caveats), the necessary exposure for a potential discovery at $5\,\sigma$ with the expected background rate would require a run length of $\unit[390]{days}$, for an expected $6.4$ axion events and $1.6$ background events.

In conclusion, early projections for the S140 show that this technology has the potential for probing the preferred parameter space of the solar KK axion model within achievable exposure levels. Furthermore, even restricted runs would set new constraints on the model. For a moderate run length of $\unit[30]{days}$, and taking still the $20\%$ background-to-axion ratio, the projected excluded axion-photon coupling strength is $g_{a\gamma\gamma}<\unit[2.09\cdot 10^{-13}]{GeV^{-1}}$ for a local density of KK axions of $n_{_{\mathrm{DLZ}}}=\unit[4.07\cdot10^{13}]{m^{-3}}$, a 4.3 times stronger limit than the one set with \mbox{SEDINE} in this study\footnote{Axion decay rates vary as $g_{a\gamma\gamma}^2$, so the limit on the coupling strength only decreases with the square root of the exposure. Hence why increasing the detector volume by a factor of $11$ from \mbox{SEDINE} to S140 did not decrease the constraint on $g_{a\gamma\gamma}$ by the same factor.}.


\section{CONCLUSION} 
\label{Sec:Conclusion}

In models of higher-dimensional quantum gravity, the QCD axion gains excitations of higher mass, which accumulate around the Sun and decay into two photons, as described in Ref.~\cite{DiLella2003}. For some values of the axion-photon coupling, these axion decays could explain the solar corona heating problem, and would also lead to an integrated decay rate on Earth of approximately $\unit[0.08]{events/m^3/day}$, mainly in the $5-15 \,\mathrm{keV}$ range. The usage of gaseous detectors in the search for these decays have the unique advantage of being able to achieve excellent background rejection. The two photons from the decay travel some distance in opposite directions before interacting, producing two pulses of same amplitude shortly after one another. Pulse processing methods to select for this type of event where developed, and simulations of the detector were constructed to predict their performance. The concept was tested on a prototype SPC at Queen's through $^{55}$Fe-induced argon fluorescence, keeping an axionlike detection efficiency of $25\%$ at $\unit[5.9]{keV}$, in accordance with simulations, while achieving a background rejection of at least $99.3\%$.

A $\unit[42]{day}$ long neon data taken with the $\unit[60]{cm}$ diameter SEDINE was used to set limits on the solar KK axion model. Adapting the selection criteria to the expected radioactive contamination, the background rejection was improved to $99.99\%$ in the $2-\unit[22]{keV}$, with a sensitivity to axion decays of $16.34\%$. With a single candidate event left, consistent with background expectations, \mbox{NEWS-G} sets a world-leading exclusion limit on solar KK axions, $g_{a\gamma\gamma}<\unit[8.99\cdot10^{-13}]{GeV^{-1}}$ for a KK axion density on Earth of $n_{DLZ}=\unit[4.07\cdot10^{13}]{m^{-3}}$ and two extra dimensions of size $R = \unit[1]{eV^{-1}}$. Despite the limited exposure, this is still five times stronger than the only other preexisting contraint on axion decays on Earth, previously set by XMASS with one year of data. It is also the first limit based on the search for the double-photon signature of such decays, with decay rates on Earth above approximately $\unit[2]{events/day/m^3}$ being excluded at $90\%$ C.L. for any non-relativistic particle in the $\unit[9-14]{keV}$ mass range.

The level of radioactive contamination is not low enough for SEDINE to improve this limit through increased exposure. However, the upcoming S140 detector at \mbox{SNOLAB} will benefit from higher radiopurity, in particular thanks to an electroplated copper layer on its inner surface. This will lead to a ratio of only $20\%$ between background and axion decay after background rejection, and make it fully capable of probing the preferred parameter space of the solar KK axion model. This would require $\unit[400]{days}$ of data taking for the S140's stay at \mbox{SNOLAB}, with only $\unit[30]{days}$ being enough to improve on SEDINE's limit by a factor of $4.3$.


\section*{Acknowledgments}

The help of the technical staff of the Laboratoire Souterrain de Modane is gratefully acknowledged. The
low activity prototype operated in LSM has been partially funded by the European Commission astroparticle
program Integrated Large Infrastructures for Astroparticle Science (ILIAS) (Contract No.~R113-CT-2004- 506222). This work was undertaken, in part, thanks to funding from the Canada Research Chairs program, as well as from the French National Research Agency (No.~ANR15-CE31-0008). This project has received support from the European Union's Horizon 2020 research and innovation programme under grant agreement No.~841261 (DarkSphere), and by UKRI-STFC through grants No.~ST/V006339/1 and No.~ST/S000860/1.


\appendix

\setcounter{section}{1} 

\begin{table*}[t!]
\begin{tabular*}{\textwidth}{lllllllllll}
Mass {[}keV{]} & 2.5 & 3.5     & 4.5    & 5.5    & 6.5    & 7.5    & 8.5    & 9.5    & 10.5   & 11.5    \\
Efficiency       & 0   & 0.00063 & 0.0088 & 0.042  & 0.10   & 0.17   & 0.22   & 0.26   & 0.28   & 0.28    \\
Resolution       & NA  & 19.8\%  & 20.7\% & 20.0\% & 19.2\% & 18.4\% & 17.7\% & 17.0\% & 16.4\% & 15.9\%  \\[1mm]

\hline 

Mass {[}keV{]} &  12.5   & 13.5   & 14.5   & 15.5   & 16.5   & 17.5   & 18.5   & 19.5   & 20.5   & 21.5   \\
Efficiency        & 0.26   & 0.24   & 0.21   & 0.18   & 0.15   & 0.13   & 0.10   & 0.082  & 0.066  & 0.049  \\
Resolution       & 15.4\% & 15.1\% & 14.7\% & 14.5\% & 14.2\% & 14.0\% & 13.8\% & 13.5\% & 13.0\% & 13.3\%
\end{tabular*}
\caption{Efficiency and energy resolution of the pulse processing described in this work for double-photon decay identification depending on the mass of the decaying particle; values are computed as the average within $\unit[1]{keV}$ bins. The resolution is extrapolated through simulations from the $22\%$ resolution of the $^{37}$Ar decay.}
\label{Tab:GenericLimit}
\end{table*}

\setcounter{section}{0} 
\section{CONSTRAINTS ON OTHER MODELS}
\label{App:GenericLimit}

\begin{figure}
\centering
  \includegraphics[width=\linewidth]{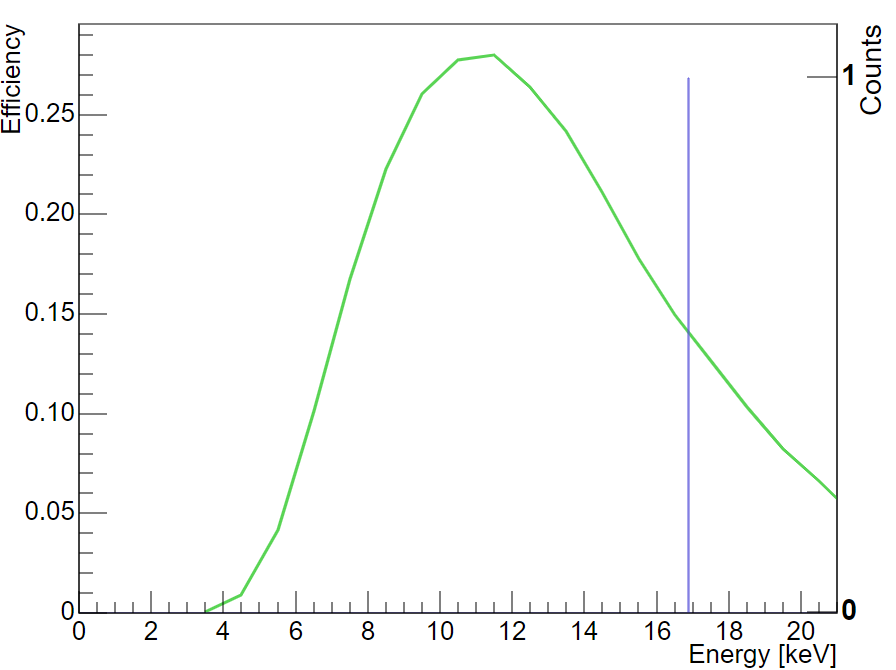}
  \caption{Detector efficiency (green, left axis) and detected events in the SEDINE detector (blue, right axis), for a total exposure of $4.3\,\mathrm{day\cdot m^3}$. Energies are corrected for attachment; the only measured event is at $\unit[16.9]{keV}$.}
  \label{fig:SEDINE_GenericLimit}
\end{figure}

The results of this search are presented in a model-independent manner in Fig.~\ref{fig:SEDINE_GenericLimit}. They can be used to set constraints on any model of non-relativistic particles decaying into two photons of same energy. To do so, the energy spectrum of decays in $\mathrm{events/day/m^3}$ should be multiplied by the efficiency curve then convolved by gaussians of relative standard deviation as given by Table~\ref{Tab:GenericLimit}. The resulting spectrum should be integrated to get the final event rate in $\mathrm{events/day/m^3}$ observed by the SEDINE detector; a subrange of energies may be chosen for integration if the signal does not cover the full $\unit[2-22]{keV}$ range. This rate is then multiplied by the $4.3\,\mathrm{day\cdot m^3}$ exposure of the SEDINE run to obtain the number of events predicted by the model. This is then compared to the number of observed events (only one at $\unit[16.9]{keV}$) as was done in Sec.~\ref{Sec:ExclLimit} to determine whether the model is excluded or not.

As an example of interpretation, any signal in the $9-14\,\mathrm{keV}$ range generated by non-relativistic particles decaying into two photons of same energy is excluded at $90\%$ C.L. for integrated decay rates higher than approximately $2\,\mathrm{events/day/m^3}$. The numbers are provided without uncertainties, although the final $24\%$ uncertainty on the final integrated efficiency for solar KK axions (see Sec.~\ref{Sec:ExclLimit}) can be used as an estimate. Precise calculations for specific models or extensions to higher energies can be performed if requested.

\bibliographystyle{elsarticle-num} 
\bibliography{AxionPaper-Biblio_truncated.bib}

\begin{thebibliography}{10}
\expandafter\ifx\csname url\endcsname\relax
  \def\url#1{\texttt{#1}}\fi
\expandafter\ifx\csname urlprefix\endcsname\relax\def\urlprefix{URL }\fi
\expandafter\ifx\csname href\endcsname\relax
  \def\href#1#2{#2} \def\path#1{#1}\fi

\bibitem{Peccei1977}
R.~D. Peccei, H.~R. Quinn, {CP} conservation in the presence of
  pseudoparticles, Phys. Rev. Lett. 38 (1977).
\newblock \href {https://doi.org/10.1103/PhysRevLett.38.1440}
  {\path{doi:10.1103/PhysRevLett.38.1440}}.

\bibitem{Kim1979}
J.~E. Kim, Weak-interaction singlet and strong cp invariance, Phys. Rev. Lett.
  43 (1979).
\newblock \href {https://doi.org/10.1103/PhysRevLett.43.103}
  {\path{doi:10.1103/PhysRevLett.43.103}}.

\bibitem{Shifman1980}
M.~A. Shifman, A.~I. Vainshtein, V.~I. Zakharov, Can confinement ensure natural
  cp invariance of strong interactions?, Nucl. Phys. B 166 (1980).
\newblock \href {https://doi.org/10.1016/0550-3213(80)90209-6}
  {\path{doi:10.1016/0550-3213(80)90209-6}}.

\bibitem{Zhitnitskij1980}
A.~R. Zhitnitskij, On possible suppression of the axion-hadron interactions,
  Yad. Fiz. 31 (1980).

\bibitem{Dine1981}
M.~Dine, W.~Fischler, M.~Srednicki, A simple solution to the strong cp problem
  with a harmless axion, Phys. Lett. B 104 (1981).
\newblock \href {https://doi.org/10.1016/0370-2693(81)90590-6}
  {\path{doi:10.1016/0370-2693(81)90590-6}}.

\bibitem{Preskill1983}
J.~Preskill, M.~B. Wise, F.~Wilczek, Cosmology of the invisible axion, Phys.
  Lett. B 120 (1983).
\newblock \href {https://doi.org/10.1016/0370-2693(83)90637-8}
  {\path{doi:10.1016/0370-2693(83)90637-8}}.

\bibitem{Dine1983}
M.~Dine, W.~Fischler, The not-so-harmless axion, Phys. Lett. B 120 (1983).
\newblock \href {https://doi.org/10.1016/0370-2693(83)90639-1}
  {\path{doi:10.1016/0370-2693(83)90639-1}}.

\bibitem{Abbott1983}
L.~F. Abbott, P.~Sikivie, A cosmological bound on the invisible axion, Phys.
  Lett. B 120 (1983).
\newblock \href {https://doi.org/10.1016/0370-2693(83)90638-X}
  {\path{doi:10.1016/0370-2693(83)90638-X}}.

\bibitem{PDG_axion}
{Particle Data Group}, The review of particle physics, Phys. Rev. D 98~(3)
  (2019).
\newblock \href {https://doi.org/10.1103/PhysRevD.98.030001}
  {\path{doi:10.1103/PhysRevD.98.030001}}.

\bibitem{GrilliDiCortona2016}
G.~{Grilli di Cortona}, E.~Hardy, J.~{Pardo Vega}, G.~Villadoro, The {QCD}
  axion, precisely, J. of High Energy Phys. 2016 (2016).
\newblock \href {https://doi.org/10.1007/JHEP01(2016)034}
  {\path{doi:10.1007/JHEP01(2016)034}}.

\bibitem{DiLuzio2020}
L.~D. Luzio, M.~Giannotti, E.~Nardi, L.~Visinelli, The landscape of {QCD} axion
  models, Phys. Rep. 870 (2020).
\newblock \href {https://doi.org/10.1016/j.physrep.2020.06.002}
  {\path{doi:10.1016/j.physrep.2020.06.002}}.

\bibitem{Dvali1999}
N.~Arkani-Hamed, S.~Dimopoulos, G.~Dvali, Phenomenology, astrophysics, and
  cosmology of theories with submillimeter dimensions and {TeV} scale quantum
  gravity, Phys. Rev. D 59 (1999).
\newblock \href {https://doi.org/10.1103/PhysRevD.59.086004}
  {\path{doi:10.1103/PhysRevD.59.086004}}.

\bibitem{Dienes2000}
K.~R. Dienes, E.~Dudas, T.~Gherghetta, Invisible axions and large-radius
  compactifications, Phys. Rev. D 62 (2000).
\newblock \href {https://doi.org/10.1103/physrevd.62.105023}
  {\path{doi:10.1103/physrevd.62.105023}}.

\bibitem{Chang2000}
S.~Chang, S.~Tazawa, M.~Yamaguchi, Axion model in extra dimensions with {TeV}
  scale gravity, Phys. Rev. D 61 (2000).
\newblock \href {https://doi.org/10.1103/PhysRevD.61.084005}
  {\path{doi:10.1103/PhysRevD.61.084005}}.

\bibitem{DiLella2000}
L.~DiLella, A.~Pilaftsis, G.~Raffelt, K.~Zioutas, Search for solar
  {Kaluza}-{Klein} axions in theories of low-scale quantum gravity, Phys. Rev.
  D 62 (2000).
\newblock \href {https://doi.org/10.1103/PhysRevD.62.125011}
  {\path{doi:10.1103/PhysRevD.62.125011}}.

\bibitem{Primakoff1951}
H.~Primakoff, Photo-production of neutral mesons in nuclear electric fields and
  the mean life of the neutral meson, Phys. Rev. 81 (1951).
\newblock \href {https://doi.org/10.1103/PhysRev.81.899}
  {\path{doi:10.1103/PhysRev.81.899}}.

\bibitem{Raffelt1986}
G.~G. Raffelt, Astrophysical axion bounds diminished by screening effects,
  Phys. Rev. D 33 (2 1986).
\newblock \href {https://doi.org/10.1103/PhysRevD.33.897}
  {\path{doi:10.1103/PhysRevD.33.897}}.

\bibitem{DiLella2003}
L.~DiLella, K.~Zioutas, Observational evidence for gravitationally trapped
  massive axion(-like) particles, Astropart. Phys. 19 (2003).
\newblock \href {https://doi.org/10.1016/S0927-6505(02)00186-X}
  {\path{doi:10.1016/S0927-6505(02)00186-X}}.

\bibitem{Barth2013}
K.~Barth, et~al., Cast constraints on the axion-electron coupling, J. Cosmol.
  Astropart. Phys. 2013 (5 2013).
\newblock \href {https://doi.org/10.1088/1475-7516/2013/05/010}
  {\path{doi:10.1088/1475-7516/2013/05/010}}.

\bibitem{Redondo2013}
J.~Redondo, Solar axion flux from the axion-electron coupling, J. Cosmol.
  Astropart. Phys. 2013 (12 2013).
\newblock \href {https://doi.org/10.1088/1475-7516/2013/12/008}
  {\path{doi:10.1088/1475-7516/2013/12/008}}.

\bibitem{Grotrian1939}
W.~Grotrian, Zur frage der deutung der linien im spektrum der sonnenkorona,
  Naturwissenschaften 27 (3 1939).
\newblock \href {https://doi.org/10.1007/BF01488890}
  {\path{doi:10.1007/BF01488890}}.

\bibitem{Edlen1942}
B.~Edlen, Die deutung der emissionslinien im spektrum der sonnenkorona, Z.
  Phys. 22 (1942).

\bibitem{Parnell2012}
C.~E. Parnell, I.~{De Moortel}, A contemporary view of coronal heating, Phil.
  Trans. R. Soc. A 370 (2012).
\newblock \href {https://doi.org/10.1098/rsta.2012.0113}
  {\path{doi:10.1098/rsta.2012.0113}}.

\bibitem{Cranmer2015}
S.~R. Cranmer, et~al., The role of turbulence in coronal heating and solar wind
  expansion, Phil. Trans. R. Soc. A 373 (2015).
\newblock \href {https://doi.org/10.1098/rsta.2014.0148}
  {\path{doi:10.1098/rsta.2014.0148}}.

\bibitem{DeMoortel2015}
I.~{De Moortel}, P.~Browning, Recent advances in coronal heating, Phil. Trans.
  R. Soc. A 373 (2015).
\newblock \href {https://doi.org/10.1098/rsta.2014.0269}
  {\path{doi:10.1098/rsta.2014.0269}}.

\bibitem{Sakurai2017}
T.~Sakurai, Heating mechanisms of the solar corona, Proc. Jpn. Acad. Ser. B 93
  (2017).
\newblock \href {https://doi.org/10.2183/pjab.93.006}
  {\path{doi:10.2183/pjab.93.006}}.

\bibitem{Klimchuk2006}
J.~A. Klimchuk, On solving the coronal heating problem, Sol. Phys. 234 (2006).
\newblock \href {https://doi.org/10.1007/s11207-006-0055-z}
  {\path{doi:10.1007/s11207-006-0055-z}}.

\bibitem{Erdelyi2007}
R.~Erdélyi, I.~Bailai, Heating of the solar and stellar coronae: A review,
  Astron. Nachr. 328 (2007).
\newblock \href {https://doi.org/10.1002/asna.200710803}
  {\path{doi:10.1002/asna.200710803}}.

\bibitem{Orlando2000}
S.~Orlando, G.~Peres, F.~Reale, The {Sun} as an {X}‐ray star. {I}. deriving
  the emission measure distribution versus temperature of the whole solar
  corona from the {Yohkoh} / soft {X}‐ray telescope data, Astrophys. J. 528
  (2000).
\newblock \href {https://doi.org/10.1086/308137} {\path{doi:10.1086/308137}}.

\bibitem{Peres2000}
G.~Peres, et~al., The {Sun} as an {X}‐ray star. {II}. using the {Yohkoh} /
  soft {X}‐ray telescope–derived solar emission measure versus temperature
  to interpret stellar {X}‐ray observations, Astrophys. J. 528 (2000).
\newblock \href {https://doi.org/10.1086/308136} {\path{doi:10.1086/308136}}.

\bibitem{Morgan2005}
B.~Morgan, et~al., Searches for solar {Kaluza}-{Klein} axions with gas {TPCs},
  Astropart. Phys. 23 (2005).
\newblock \href {https://doi.org/10.1016/j.astropartphys.2005.01.002}
  {\path{doi:10.1016/j.astropartphys.2005.01.002}}.

\bibitem{Oka2017}
N.~Oka, et~al., Search for solar {Kaluza}-{Klein} axions by annual modulation
  with the {XMASS-I} detector, Prog. Theor. Exp. Phys. 2017 (2017).
\newblock \href {https://doi.org/10.1093/ptep/ptx137}
  {\path{doi:10.1093/ptep/ptx137}}.

\bibitem{Ahmad2002}
Q.~R. Ahmad, et~al., Direct evidence for neutrino flavor transformation from
  neutral-current interactions in the {Sudbury} {Neutrino} {Observatory}, Phys.
  Rev. Lett. 89 (2002).
\newblock \href {https://doi.org/10.1103/PhysRevLett.89.011301}
  {\path{doi:10.1103/PhysRevLett.89.011301}}.

\bibitem{Arnaud2018}
Q.~Arnaud, et~al., First results from the {NEWS-G} direct dark matter search
  experiment at the {LSM}, Astropart. Phys. 97 (2018).
\newblock \href {https://doi.org/10.1016/j.astropartphys.2017.10.009}
  {\path{doi:10.1016/j.astropartphys.2017.10.009}}.

\bibitem{Brossard2020Thesis}
A.~Brossard, Optimization of spherical proportional counter backgrounds and
  response for low mass dark matter search, Ph.D. thesis, Queen's University
  (2020).

\bibitem{Giomataris2008}
I.~Giomataris, et~al., A novel large-volume spherical detector with
  proportional amplification read-out, JINST 3 (2008).
\newblock \href {https://doi.org/10.1088/1748-0221/3/09/P09007}
  {\path{doi:10.1088/1748-0221/3/09/P09007}}.

\bibitem{Piquemal2012}
F.~Piquemal, Modane underground laboratory: Status and project, Eur. Phys. J.
  Plus 127 (2012).
\newblock \href {https://doi.org/10.1140/epjp/i2012-12110-3}
  {\path{doi:10.1140/epjp/i2012-12110-3}}.

\bibitem{Laubenstein2004}
M.~Laubenstein, et~al., Underground measurements of radioactivity, Appl.
  Radiat. Isot. 61 (2004).
\newblock \href {https://doi.org/10.1016/j.apradiso.2004.03.039}
  {\path{doi:10.1016/j.apradiso.2004.03.039}}.

\bibitem{Bougamont2017}
E.~Bougamont, et~al., Neutron spectroscopy with the spherical proportional
  counter based on nitrogen gas, Nucl. Instrum. Methods Phys. Res. A 847
  (2017).
\newblock \href {https://doi.org/10.1016/j.nima.2016.11.007}
  {\path{doi:10.1016/j.nima.2016.11.007}}.

\bibitem{Savvidis2018}
I.~Savvidis, et~al., Low energy recoil detection with a spherical proportional
  counter, Nucl. Instrum. Methods Phys. Res. A 877 (1 2018).
\newblock \href {https://doi.org/10.1016/j.nima.2017.09.014}
  {\path{doi:10.1016/j.nima.2017.09.014}}.

\bibitem{Meregaglia2019}
A.~Meregaglia, A new neutrinoless double beta decay experiment: {R2D2}, J.
  Phys. Conf. Ser. 1312 (9 2019).
\newblock \href {https://doi.org/10.1088/1742-6596/1312/1/012002}
  {\path{doi:10.1088/1742-6596/1312/1/012002}}.

\bibitem{NIST}
J.~H. Hubbell, S.~M. Seltzer,
  \href{https://physics.nist.gov/PhysRefData/XrayMassCoef/tab3.html}{Tables of
  {X}-ray mass attenuation coefficients and mass energy-absorption coefficients
  (version 1.4)} (2004).
\newblock \href {https://doi.org/10.18434/T4D01F} {\path{doi:10.18434/T4D01F}}.
\newline\urlprefix\url{https://physics.nist.gov/PhysRefData/XrayMassCoef/tab3.html}

\bibitem{NIST_report}
P.~D. Higgins, et~al., Mass energy-transfer and mass energy-absorption
  coefficients, including in-flight positron annihilation for photon energies 1
  {keV} to {100 MeV}, Tech. Rep. PB92-126473/XAB, NISTIR-4680, NIST (NML)
  (1991).

\bibitem{Shockley1938}
W.~Shockley, Currents to conductors induced by a moving point charge, J. Appl.
  Phys. 9 (1938).
\newblock \href {https://doi.org/10.1063/1.1710367}
  {\path{doi:10.1063/1.1710367}}.

\bibitem{He2001}
Z.~He, Review of the shockley-ramo theorem and its application in semiconductor
  gamma-ray detectors, Nucl. Instrum. Methods Phys. Res. A 463 (2001).
\newblock \href {https://doi.org/10.1016/S0168-9002(01)00223-6}
  {\path{doi:10.1016/S0168-9002(01)00223-6}}.

\bibitem{Vazquez2020Thesis}
F.~A. Vazquez~de Sola~Fernandez, Solar {KK} axion search with {NEWS-G}, Ph.D.
  thesis, Queen's University (2020).

\bibitem{NIST_STAR}
M.~J. Berger, J.~S. Coursey, M.~A. Zucker, J.~Chang,
  \href{https://physics.nist.gov/PhysRefData/Star/Text/ESTAR.html}{{ESTAR},
  {PSTAR}, and {ASTAR}: Computer programs for calculating stopping-power and
  range tables for electrons, protons, and helium ions (version 2.0.1)} (2017).
\newblock \href {https://doi.org/10.18434/T4NC7P} {\path{doi:10.18434/T4NC7P}}.
\newline\urlprefix\url{https://physics.nist.gov/PhysRefData/Star/Text/ESTAR.html}

\bibitem{Agostinelli2003}
S.~Agostinelli, et~al., {GEANT4} - a simulation toolkit, Nucl. Instrum. Methods
  Phys. Res. A 506 (2003).
\newblock \href {https://doi.org/10.1016/S0168-9002(03)01368-8}
  {\path{doi:10.1016/S0168-9002(03)01368-8}}.

\bibitem{Allison2006}
J.~Allison, et~al., {Geant4} developments and applications, IEEE Trans. Nucl.
  Sci. 53 (2006).
\newblock \href {https://doi.org/10.1109/TNS.2006.869826}
  {\path{doi:10.1109/TNS.2006.869826}}.

\bibitem{Allison2016}
J.~Allison, et~al., Recent developments in {GEANT4}, Nucl. Instrum. Methods
  Phys. Res. A 835 (2016).
\newblock \href {https://doi.org/10.1016/j.nima.2016.06.125}
  {\path{doi:10.1016/j.nima.2016.06.125}}.

\bibitem{Inokuti1975}
M.~Inokuti, Ionization yields in gases under electron irradiation, Radiat. Res.
  64 (1975).
\newblock \href {https://doi.org/10.2307/3574165} {\path{doi:10.2307/3574165}}.

\bibitem{Doke1992}
T.~Doke, N.~Ishida, M.~Kase, Fano factors in rare gases and their mixtures,
  Nucl. Instrum. Methods Phys. Res. B 63 (3 1992).
\newblock \href {https://doi.org/10.1016/0168-583X(92)95207-8}
  {\path{doi:10.1016/0168-583X(92)95207-8}}.

\bibitem{Grosswendt1984}
B.~Grosswendt, Statistical fluctuations of the ionisation yield of low-energy
  electrons in {He}, {Ne} and {Ar}, J. Phys. B 17 (1984).
\newblock \href {https://doi.org/10.1088/0022-3700/17/7/022}
  {\path{doi:10.1088/0022-3700/17/7/022}}.

\bibitem{Santos1999}
F.~P. Santos, et~al., {W}-value and the {F}ano factor for x-rays in xenon-neon
  gas mixtures: a {M}onte {C}arlo simulation study, IEEE Nucl. Sci. Symp. Med.
  Imaging Conf. 1 (1999).
\newblock \href {https://doi.org/10.1109/nssmic.1998.775206}
  {\path{doi:10.1109/nssmic.1998.775206}}.

\bibitem{Durnford2018}
D.~Durnford, Q.~Arnaud, G.~Gerbier, Novel approach to assess the impact of the
  fano factor on the sensitivity of low-mass dark matter experiments, Phys.
  Rev. D 98 (2018).
\newblock \href {https://doi.org/10.1103/PhysRevD.98.103013}
  {\path{doi:10.1103/PhysRevD.98.103013}}.

\bibitem{COMSOL}
I.~COMSOL, \href{https://www.comsol.com}{{COMSOL} {Multiphysics} reference
  manual (version 5.3)}.
\newline\urlprefix\url{https://www.comsol.com}

\bibitem{Biagi1999}
S.~F. Biagi, {Monte} {Carlo} simulation of electron drift and diffusion in
  counting gases under the influence of electric and magnetic fields, Nucl.
  Instrum. Methods Phys. Res. A 421 (1999).
\newblock \href {https://doi.org/10.1016/S0168-9002(98)01233-9}
  {\path{doi:10.1016/S0168-9002(98)01233-9}}.

\bibitem{Alkhazov1970}
G.~D. Alkhazov, Statistics of electron avalanches and ultimate resolution of
  proportional counters, Nucl. Instrum. Methods 89 (1970).
\newblock \href {https://doi.org/10.1016/0029-554X(70)90818-9}
  {\path{doi:10.1016/0029-554X(70)90818-9}}.

\bibitem{Arnaud2019}
Q.~Arnaud, et~al., Precision laser-based measurements of the single electron
  response of spherical proportional counters for the {NEWS-G} light dark
  matter search experiment, Phys. Rev. D 99 (2019).
\newblock \href {https://doi.org/10.1103/PhysRevD.99.102003}
  {\path{doi:10.1103/PhysRevD.99.102003}}.

\bibitem{TableRadionuclides}
M.-M. Bé, et~al., Table of Radionuclides, Vol.~3 of Monographie BIPM-5, Bureau
  International des Poids et Mesures, 2006.

\bibitem{TableRadioactiveIsotopes}
S.~Y.~F. Chu, L.~P. Ekstrom, R.~B. Firestone,
  \href{http://nucleardata.nuclear.lu.se/toi/}{{WWW} table of radioactive
  isotopes} (1999).
\newline\urlprefix\url{http://nucleardata.nuclear.lu.se/toi/}

\bibitem{Bambynek1972}
W.~Bambynek, et~al., {X}-ray fluorescence yields, {Auger}, and
  {Coster}-{Kronig} transition probabilities, Rev. Mod. Phys. 44 (1972).
\newblock \href {https://doi.org/10.1103/RevModPhys.44.716}
  {\path{doi:10.1103/RevModPhys.44.716}}.

\bibitem{Krause1979}
M.~O. Krause, Atomic radiative and radiationless yields for {K} and {L} shells,
  J. Phys. Chem. Ref. Data 8 (1979).
\newblock \href {https://doi.org/10.1063/1.555594}
  {\path{doi:10.1063/1.555594}}.

\bibitem{Hubbell1994}
J.~H. Hubbell, et~al., A review, bibliography, and tabulation of {K}, {L}, and
  higher atomic shell {X}-ray fluorescence yields, J. Phys. Chem. Ref. Data 23
  (1994).
\newblock \href {https://doi.org/10.1063/1.555955}
  {\path{doi:10.1063/1.555955}}.

\bibitem{Shukla2018}
P.~Shukla, S.~Sankrith, Energy and angular distributions of atmospheric muons
  at the {Earth}, Int. J. Mod. Phys. A 33 (2018).
\newblock \href {https://doi.org/10.1142/S0217751X18501750}
  {\path{doi:10.1142/S0217751X18501750}}.

\bibitem{SauliLectures}
F.~Sauli, Principles of operation of multiwire proportional and drift chambers,
  Tech. Rep. CERN-77-09, CERN (1977).

\bibitem{Skullerud1990}
H.~R. Skullerud, P.~H. Larsen, Mobility and diffusion of atomic helium and neon
  ions in their parent gases, J. Phys. B 23 (1990).
\newblock \href {https://doi.org/10.1088/0953-4075/23/6/010}
  {\path{doi:10.1088/0953-4075/23/6/010}}.

\bibitem{Hornbeck1951}
J.~A. Hornbeck, The drift velocities of molecular and atomic ions in helium,
  neon, and argon, Phys. Rev. 84 (1951).
\newblock \href {https://doi.org/10.1103/PhysRev.84.615}
  {\path{doi:10.1103/PhysRev.84.615}}.

\bibitem{Agnese2013}
R.~Agnese, et~al., Demonstration of surface electron rejection with interleaved
  germanium detectors for dark matter searches, Appl. Phys. Lett. 103 (2013).
\newblock \href {https://doi.org/10.1063/1.4826093}
  {\path{doi:10.1063/1.4826093}}.

\bibitem{GARWOOD1936}
F.~GARWOOD, Fiducial limits for the {Poisson} distribution, Biometrika 28
  (1936).
\newblock \href {https://doi.org/10.1093/biomet/28.3-4.437}
  {\path{doi:10.1093/biomet/28.3-4.437}}.

\bibitem{Anastassopoulos2017}
V.~Anastassopoulos, et~al., New {CAST} limit on the axion-photon interaction,
  Nat. Phys. 13 (2017).
\newblock \href {https://doi.org/10.1038/nphys4109}
  {\path{doi:10.1038/nphys4109}}.

\bibitem{Duncan2010}
F.~Duncan, A.~J. Noble, D.~Sinclair, The construction and anticipated science
  of {SNOLAB}, Annu. Rev. Nucl. Part. Sci. 60 (2010).
\newblock \href {https://doi.org/10.1146/annurev.nucl.012809.104513}
  {\path{doi:10.1146/annurev.nucl.012809.104513}}.

\bibitem{Giganon2017}
A.~Giganon, et~al., A multiball read-out for the spherical proportional
  counter, JINST 12 (2017).
\newblock \href {https://doi.org/10.1088/1748-0221/12/12/P12031}
  {\path{doi:10.1088/1748-0221/12/12/P12031}}.

\bibitem{Giomataris2020}
I.~Giomataris, et~al., A resistive achinos multi-anode structure with dlc
  coating for spherical proportional counters, JINST 15 (2020).
\newblock \href {https://doi.org/10.1088/1748-0221/15/11/P11023}
  {\path{doi:10.1088/1748-0221/15/11/P11023}}.

\bibitem{Balogh2021}
L.~Balogh, et~al., Copper electroplating for background suppression in the
  {NEWS-G} experiment, Nucl. Instrum. Methods Phys. Res. A 988 (2021).
\newblock \href {https://doi.org/https://doi.org/10.1016/j.nima.2020.164844}
  {\path{doi:https://doi.org/10.1016/j.nima.2020.164844}}.

\bibitem{Giroux2019}
G.~Giroux, P.~Gros, I.~Katsioulas, The search for light dark matter with the
  {NEWS-G} spherical proportional counter, J. Phys. Conf. Ser. 1312 (2019).
\newblock \href {https://doi.org/10.1088/1742-6596/1312/1/012008}
  {\path{doi:10.1088/1742-6596/1312/1/012008}}.

\end{thebibliography}



\end{document}